\documentclass{cDSS2e}

\usepackage{amsfonts}
\usepackage{amsmath}
\usepackage{xspace}
\usepackage{color,graphicx}
\usepackage{epsf}
\usepackage{epsfig}
\usepackage{psfrag}
\usepackage{subfigure}

\graphicspath{{../figures/}{figures/}}

\usepackage{cite}

\newcommand{\Rset}{\mathbb{R}}

\newcommand{\HAin}{{\mathbf{H}_A^{\rm in}}}
\newcommand{\HAout}{{\mathbf{H}_A^{\rm out}}}
\newcommand{\HBin}{{\mathbf{H}_B^{\rm in}}}
\newcommand{\HBout}{{\mathbf{H}_B^{\rm out}}}
\newcommand{\HCin}{{\mathbf{H}_C^{\rm in}}}
\newcommand{\HCout}{{\mathbf{H}_C^{\rm out}}}

\newcommand{\DAmin}{{D_A^{\rm min}}}
\newcommand{\DAmax}{{D_A^{\rm max}}}
\newcommand{\DBmin}{{D_B^{\rm min}}}
\newcommand{\DBmax}{{D_B^{\rm max}}}
\newcommand{\DCmin}{{D_C^{\rm min}}}
\newcommand{\DCmax}{{D_C^{\rm max}}}
\newcommand{\DABCmin}{{D_{ABC}^{\rm min}}}
\newcommand{\DABCmax}{{D_{ABC}^{\rm max}}}
\newcommand{\DCAmin}{{D_{CA}^{\rm min}}}
\newcommand{\DCAmax}{{D_{CA}^{\rm max}}}
\newcommand{\DXone}{{D_{X,1}}}
\newcommand{\DXtwo}{{D_{X,2}}}
\newcommand{\DXmax}{{D_{X}^{\rm max}}}
\newcommand{\DXmin}{{D_{X}^{\rm min}}}

\newcommand{\thetaAout}{{\theta_3^{(A_{\rm out})}}}
\newcommand{\riiiBin}{{r_3^{(B_{\rm in})}}}
\newcommand{\thetaBin}{{\theta_3^{(B_{\rm in})}}}

\newcommand{\xiBout}{{x_1^{(B_{\rm out})}}}
\newcommand{\thetaBout}{{\theta_3^{(B_{\rm out})}}}
\newcommand{\xiCin}{{x_1^{(C_{\rm in})}}}
\newcommand{\thetaCin}{{\theta_3^{(C_{\rm in})}}}

\newcommand{\xiiCout}{{x_2^{(C_{\rm out})}}}
\newcommand{\thetaCout}{{\theta_3^{(C_{\rm out})}}}
\newcommand{\xiiAin}{{x_2^{(A_{\rm in})}}}
\newcommand{\thetaAin}{{\theta_3^{(A_{\rm in})}}}

\newcommand{\rA}{{r_A}}
\newcommand{\cA}{{c_A}}
\newcommand{\eA}{{e_A}}

\newcommand{\rB}{{r_B}}
\newcommand{\cB}{{c_B}}
\newcommand{\eBx}{{e_{Bx}}}
\newcommand{\eBy}{{e_{By}}}

\newcommand{\rC}{{r_C}}
\newcommand{\cC}{{c_C}}
\newcommand{\eC}{{e_C}}

\newcommand{\deltaA}{{\delta_A}}
\newcommand{\deltaBx}{{\delta_{Bx}}}
\newcommand{\deltaBy}{{\delta_{By}}}
\newcommand{\deltaCX}{{\delta_{CX}}}
\newcommand{\deltaCY}{{\delta_{CY}}}
\newcommand{\deltaX}{{\delta_{X}}}
\newcommand{\deltaY}{{\delta_{Y}}}
\newcommand{\deltaCP}{{\delta_{CP}}}
\newcommand{\deltaCQ}{{\delta_{CQ}}}
\newcommand{\deltaBmax}{{\delta_{B}^{\rm max}}}
\newcommand{\deltaBmin}{{\delta_{B}^{\rm min}}}
\newcommand{\deltaCmax}{{\delta_{C}^{\rm max}}}
\newcommand{\deltaCmin}{{\delta_{C}^{\rm min}}}
\newcommand{\deltamax}{{\delta^{\rm max}}}
\newcommand{\deltamin}{{\delta^{\rm min}}}

\begin{document}
\doi{}
 \issn{}
\issnp{00} \jvol{00} \jnum{00} \jyear{2009} \jmonth{March}
\markboth{}{}

\title{A mechanism for switching near a heteroclinic network}

\author{Vivien Kirk,\thanks{v.kirk@auckland.ac.nz}
Department of Mathematics, University of Auckland,
\break Private Bag 92019, Auckland, New Zealand
\break Emily Lane,\thanks{e.lane@niwa.co.nz}
National Institute of Water and Atmospheric Research,
\break P O Box 8602, Christchurch, New Zealand
\break Claire M. Postlethwaite, \thanks{c.postlethwaite@math.auckland.ac.nz}
Department of Mathematics, University of Auckland,
\break Private Bag 92019, Auckland, New Zealand
\break Alastair M. Rucklidge\thanks{a.m.rucklidge@leeds.ac.uk},
Department of Applied Mathematics, University of Leeds,
\break       Leeds LS2 9JT, UK
\break Mary Silber,\thanks{m-silber@northwestern.edu}
Department of Engineering Sciences and Applied Mathematics,
Northwestern University, Evanston, IL 60208, USA}

\received{\today}
\maketitle

\begin{abstract}
We describe an example of a structurally stable heteroclinic network
for which nearby orbits exhibit irregular but sustained switching
between the various sub-cycles in the network. The mechanism for switching
is the presence of spiralling due to complex eigenvalues in the
flow linearised about one of the equilibria common to all cycles in the
network. We construct and use return maps to investigate the asymptotic
stability of the network, and show that switching is ubiquitous near
the network.
Some of the unstable manifolds involved in the network are two-dimensional;
we develop a technique to account for all trajectories on those manifolds.
A simple numerical example illustrates the
rich dynamics that can result from the interplay between
the various cycles in the network.
 \end{abstract}

% possible 2000 Mathematics Subject Classification (lifted from similar papers)
% 37C10, 37C29, 37C80, 37D45, 37C50, 34F05
% primary: 37G30; secondary: 37C10, 34C37, 37C29, 34C28, 37C80
% 34C37, 58F14

% Keywords: heteroclinic network; switching; heteroclinic cycles; anything else...?

\section{Introduction}
\label{sec:intro}

Heteroclinic cycles and networks are invariant sets that can occur in
a structurally stable way in systems with symmetry, and are known to
provide a robust mechanism for intermittent behaviour in these
systems.
For the purposes of this paper, we adopt the following definitions
of heteroclinic cycles and heteroclinic networks.
In the literature, there are more complicated definitions~\cite{Ashwin1999d},
but the
simpler definitions presented here suffice for our purposes.
For a finite-dimensional system of
ordinary differential equations (ODEs), we define:

{\bf Definition.}
A {\it heteroclinic cycle} $\mathcal{C}$ is a finite collection of equilibria
$\{\xi_1, \dots , \xi_n\}$ of the ODEs, together with a set of heteroclinic connections
$\{\gamma_1(t), \dots , \gamma_n(t) \}$, where $\gamma_j(t)$ is a solution of
the ODEs such that $\gamma_j(t) \to \xi_j$
as $t \to -\infty$ and $\gamma_j(t) \to \xi_{j+1}$
as $t \to \infty$, and where $\xi_{n+1} \equiv \xi_1$.

{\bf Definition.}
Let ${\mathcal{C}_1, \mathcal{C}_2, \dots }$ be a collection of two or more
heteroclinic cycles.
We say that ${\mathcal N} = \bigcup_i \mathcal{C}_i$
forms a {\it heteroclinic network} if
for each pair of equilibria in the network, there is a sequence of heteroclinic
connections joining the equilibria. That is, for any $\xi_j,
\xi_k\in\mathcal{N}$, we can find a sequence of heteroclinic connections
$\{\gamma_{p_1}(t),\dots,\gamma_{p_l}(t)\}\in\mathcal{N}$ and a sequence of equilibria
$\{\xi_{m_1},\dots,\xi_{m_{l+1}}\}\in\mathcal{N}$ such that
$\xi_{m_1}\equiv\xi_j$, $\xi_{m_{l+1}}\equiv \xi_k$ and $\gamma_{p_i}$ is a
heteroclinic connection between $\xi_{m_i}$ and $\xi_{m_{i+1}}$.

Under this definition, a heteroclinic network is a connected collection of
heteroclinic cycles, possibly infinite in number.
We allow for an infinite
number of cycles to co-exist in a network, as can occur, for instance, when one
of the equilibria has a two-dimensional unstable manifold and
there is a continuum of heteroclinic connections between that equilibrium
and another. However, we restrict to the case where the set of all equilibria in the network
is finite.
In general,  heteroclinic orbits can connect invariant sets other
than equilibria, such as periodic orbits or chaotic saddles; we do not consider
this possibility here.

Structurally stable heteroclinic cycles in symmetric systems have been studied
extensively in recent years, with a canonical example arising in the context of
rotating Rayleigh--B\'enard convection~\cite{Busse1980a} being analyzed
in~\cite{Guckenheimer1988}. A good deal is now known about conditions for
existence and stability of heteroclinic cycles (e.g.,
\cite{Krupa1995,Krupa2004,Melbourne1991,Ashwin1998d}), and some results are also known
about bifurcations of heteroclinic cycles and networks (e.g.,
\cite{Scheel1992,Chossat1997,Postlethwaite2006,Driesse2009,Ashwin2002a}) and the effect on the dynamics
of small symmetry breaking (e.g.,
\cite{Melbourne1989a,Chossat1993,Sandstede1995,Melbourne2001,Kirk2008}) or the
addition of noise (e.g.,~\cite{Armbruster2003,Stone1990,Ashwin2004a}).  Some experimental observations of
near-heteroclinic cycles have been reported (e.g., see~\cite{Nore2005} for a
recent example); experimental noise and small symmetry-breaking effects prevent
exact heteroclinic cycles from occurring, but near-heteroclinic structures are
seen in certain regimes.

The dynamics near networks of heteroclinic cycles has been studied in, for
instance, \cite{Kirk1994,Brannath1994,Ashwin1998a,Aguiar2005,Postlethwaite2005,Driesse2009}.
There are some natural questions to ask about the dynamics near heteroclinic
networks. For example, is it the case that one cycle in the network is
dominant, in that most trajectories near the network are attracted to that
cycle, resulting in the network structure not being observed?  Can more than
one cycle be observed? Are there trajectories that switch between the cycles in
the network in a sustained way, visiting all parts of the network eventually?
Partial answers to these questions have been established.

Krupa and
Melbourne~\cite{Krupa1995a} find conditions under which one cycle dominates in
certain cases. However, their  analysis does not cover the example of interest
in this paper.

Kirk and Silber~\cite{Kirk1994} construct an example where more than one cycle
is observable, specifically showing that open sets of trajectories near the
network may be attracted to each of the two primary cycles in their network.
Despite both cycles being observable, there is no sustained switching in this
example: an orbit may switch from one cycle to the other initially but may not
switch back again. The effect of small noise on the network in~\cite{Kirk1994}
is studied in~\cite{Armbruster2003}, where it is shown that noise can either
induce switching of trajectories between cycles in the network or enhance the
attractivity of certain cycles.

Aguiar et al.~\cite{Aguiar2005} describe an example, motivated by conjectures
of Field~\cite{Field1996a}, where trajectories switch between excursions about
different cycles in a heteroclinic network. In this example some of the
connections in the network result from transversal intersections between stable
and unstable manifolds of equilibria, with the consequence that the network
probably does not attract open sets of initial conditions. Because of the
reinjection mechanisms built into the network, trajectories will make repeated
passes near the transversal intersections and there are trajectories that
follow arbitrarily complicated paths around the network, but these trajectories
will not approach the network. Another example of this type is found in
\cite{Aguiar2009a}.

Postlethwaite and Dawes~\cite{Postlethwaite2005} examined an example of a
heteroclinic network in which trajectories can exhibit periodic or aperiodic
patterns of excursions past the various cycles in the network. The mechanism
inducing switching between cycles in this network is a transverse instability
of each cycle in one direction. Almost all trajectories near a cycle eventually
leave that cycle for another cycle, but that cycle in turn is unstable in a
transverse direction and trajectories eventually leave that cycle too,
ultimately returning to a neighbourhood of the original cycle. This mechanism
operates when the network as a whole is essentially asymptotically stable, so trajectories
get closer to the network as they cycle around the network.

Ashwin et al.~\cite{Ashwin2004a} describe irregular switching near a
heteroclinic network connecting periodic orbits and chaotic saddles confined to
two invariant subspaces. In this case, the mechanism determining the switching
is said to be nonlinear, since it appears to operate in a part of phase space
well away from the invariant subspaces containing the periodic orbits and
chaotic saddles. In this example, structurally stable connections again
arise from transversal intersections of manifolds.

In this paper, we present another example in which orbits near a heteroclinic
network switch repeatedly between excursions about the different cycles in the
network. The mechanism for switching is the presence of a pair
of complex eigenvalues in the linearisation of the flow about one of the
equilibrium solutions in the network.  Unlike the examples in Aguiar et
al.~\cite{Aguiar2005,Aguiar2009a}, where complex eigenvalues also occur, the heteroclinic
connections in our network are structurally stable because of the symmetries of
the problem and are non-transversal, with the consequence that, so long as
symmetries are preserved, the network can attract open sets of initial
conditions. As discussed further below, our example can exhibit an interesting
form of switching, where the network structure is evident in the long term
dynamics even though the network is not attracting. There are similarities
between this and the switching observed in \cite{Aguiar2005,Aguiar2009a}, as
discussed below.

\begin{figure}
\begin{center}
\includegraphics[width=8cm]{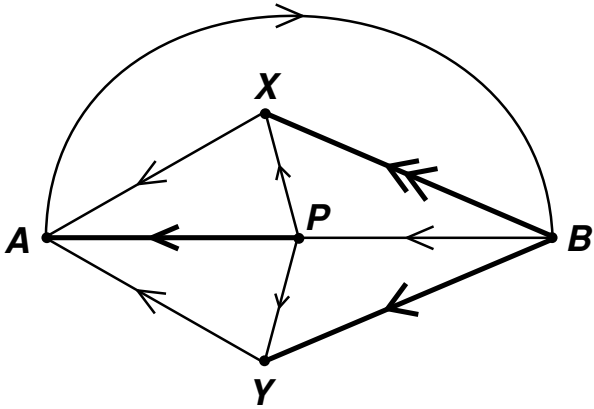}
\end{center}
\caption{Schematic diagram showing part of the
heteroclinic network studied.  For clarity, the equilibrium $Q$ is not shown; this
equilibrium plays a similar role to
equilibrium $P$ except that the one-dimensional heteroclinic connections from $Q$ connect to $-X$ and $Y$
instead of $X$ and $Y$.
The remaining (conjugate) parts of the
network are obtained under the action of the
${\mathbb Z}_2^3$  symmetry group.
The thin curves represent single heteroclinic connections while the
bold curves indicate that a two-dimensional family of connections exists between
the relevant equilibria.
The double arrowhead
on the connection from $B$ to $X$ indicates that expansion near
$B$ in the direction of this connection is stronger than the expansion
in the direction of the connection from $B$ to $Y$.}
\label{fig:network}
\end{figure}

The network we study is in
${\mathbb R}^4$ with ${\mathbb Z}_2^3$ symmetry, and is shown
schematically in Figure~\ref{fig:network}.  It consists of a set of six
equilibria denoted $A$, $B$, $X$, $Y$, $P$
and $Q$, their conjugate copies under action of the symmetries,
and the set of heteroclinic connections joining the equilibria.
Some of the heteroclinic connections occur in  two-dimensional families, as
indicated in Figure~\ref{fig:network}.
There are many different heteroclinic
cycles evident in Figure~\ref{fig:network}, e.g., $A\rightarrow
B\rightarrow X\rightarrow A$ and
$A\rightarrow B\rightarrow P\rightarrow X\rightarrow A$.
The network is the union of all these cycles, and is described in more detail
in section~\ref{sec:hetnet}.

%%%%%%%%%%%
%%%%%%%%%%%%%%

Our analysis of this example proceeds in a standard way via construction of
return maps that approximate the dynamics in a neighbourhood of the
heteroclinic network. A feature that complicates the construction is the
existence of two-dimensional unstable manifolds of some of the equilibria in
our network and hence of continua of heteroclinic connections between some
pairs of equilibria. A novel aspect of our work is the way in which we allow
for this complication;  we have developed a relatively straightforward way to keep track
of trajectories near two-dimensional unstable manifolds even when different
orbits within a manifold connect different pairs of equilibria.
The method is related to the technique used in~\cite{Rucklidge2001} to analyze
a homoclinic bifurcation in a problem with a two-dimensional unstable manifold,
and it extends
previous work on other problems with two-dimensional unstable manifolds such
as~\cite{Ashwin1998a,Ashwin1998d,Ashwin2005d,Ashwin2007,Kirk2008}.

Using these techniques, we are able to find a simple condition for asymptotic
stability (resp., instability) of the network. The condition is as expected: the product
of the ratio of contracting to expanding eigenvalues seen by a trajectory as
it traverses the network must be greater than one (resp., less than one), regardless of
the itinerary of the trajectory though the network. There is also an intermediate
case, where whether there is net contraction or expansion depends on the
itinerary of the orbit past the various equilibria in the network.
We go on to show that switching is ubiquitous in the network, whether or not
the network is asymptotically stable. If the network is asymptotically stable, then
we show that while orbits generically continue to switch as they approach the
network, visits to certain equilibria become increasingly rare.

We find that a particularly interesting form of switching can occur in our network. If one
of the cycles within the network attracts trajectories (i.e., a trajectory ends
up closer to the network after making one passage near that cycle) while other
cycles repel trajectories, then the net effect can be that a typical trajectory
approaches an attractor (possibly chaotic) that lies near the network, with the
trajectory repeatedly (but not uniformly) passing close to all parts of the
network even though the network is not itself attracting. Under this scenario
the network structure will be observed in the long term dynamics even though
the network is not attracting. We report numerical observations of this form of
switching, and defer a detailed investigation to a future paper.

The rest of this paper is organised as follows. \S\ref{sec:hetnet}~contains a
description of our heteroclinic network and details of construction of the maps
used to approximate the dynamics near the network. In~\S\ref{sec:analysis} we
find a condition for asymptotic stability of the network, and derive results
about switching
near the network. \S\ref{sec:numeg}~gives results from
numerical simulations of a system of four ordinary differential equations,
illustrating the various switching phenomena associated with our example.
Conclusions are contained in~\S\ref{sec:conc}.

\section{The heteroclinic network}
\label{sec:hetnet}

\begin{figure}
\begin{center}
\subfigure[]{\epsfig{figure=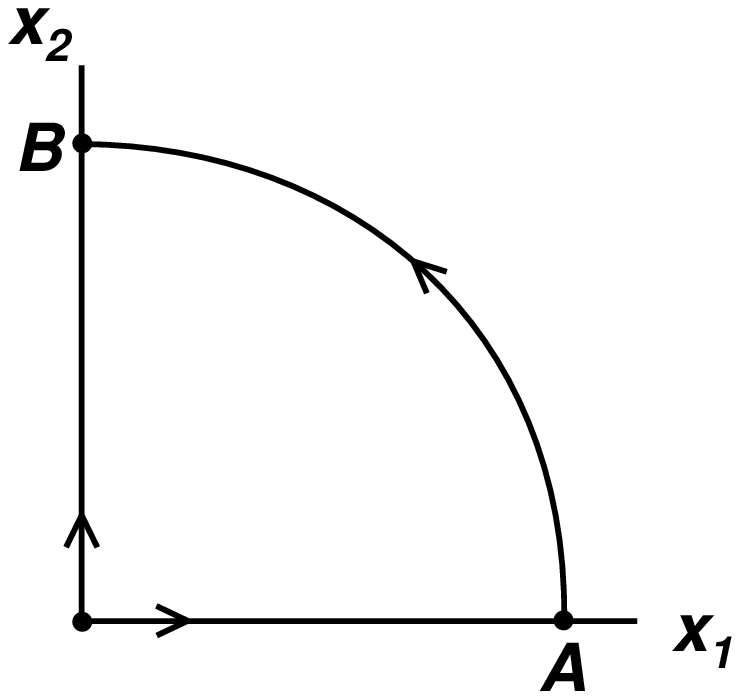,width=5cm}}\qquad
\subfigure[]{\epsfig{figure=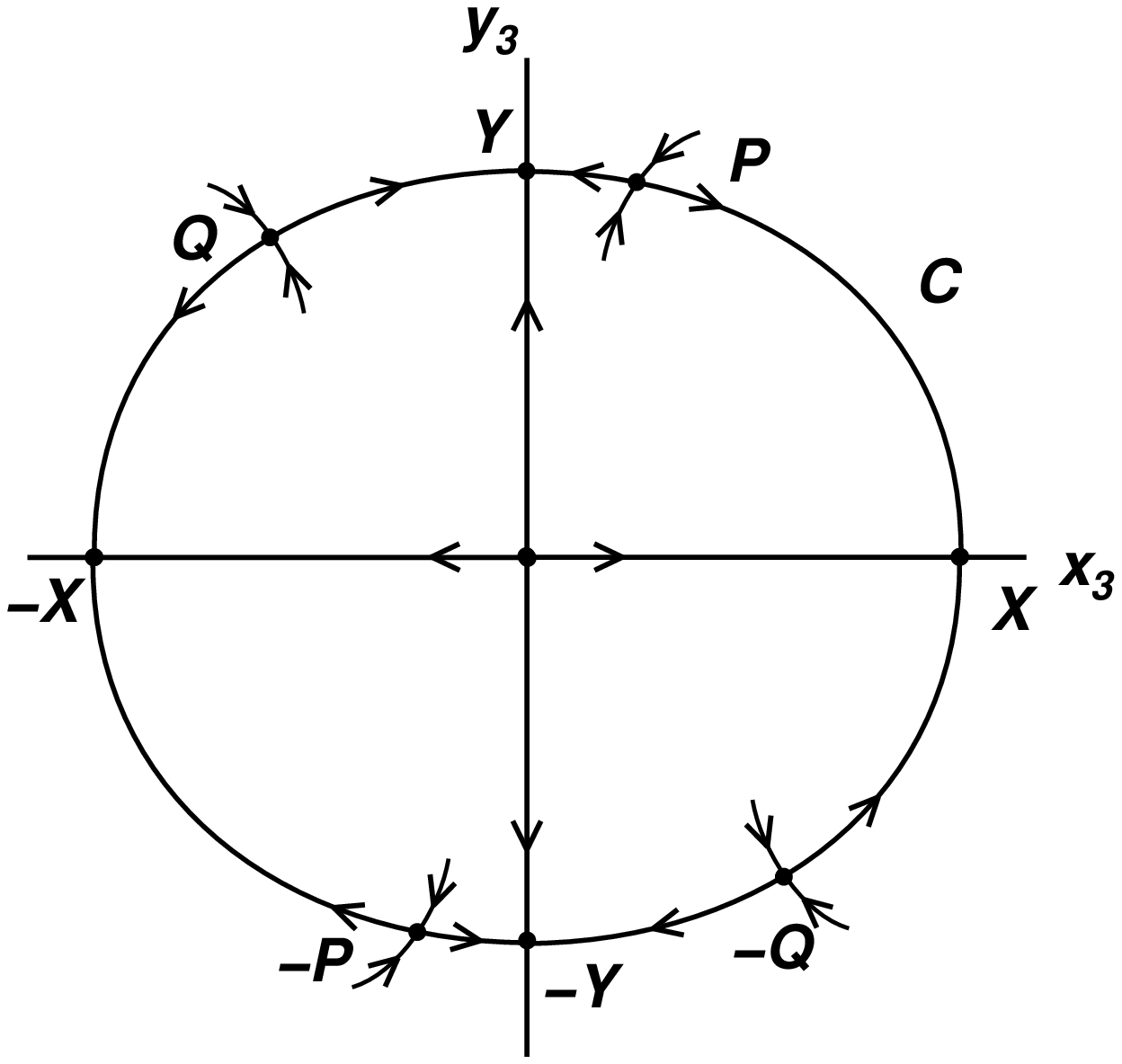,width=5cm}}
\subfigure[]{\epsfig{figure=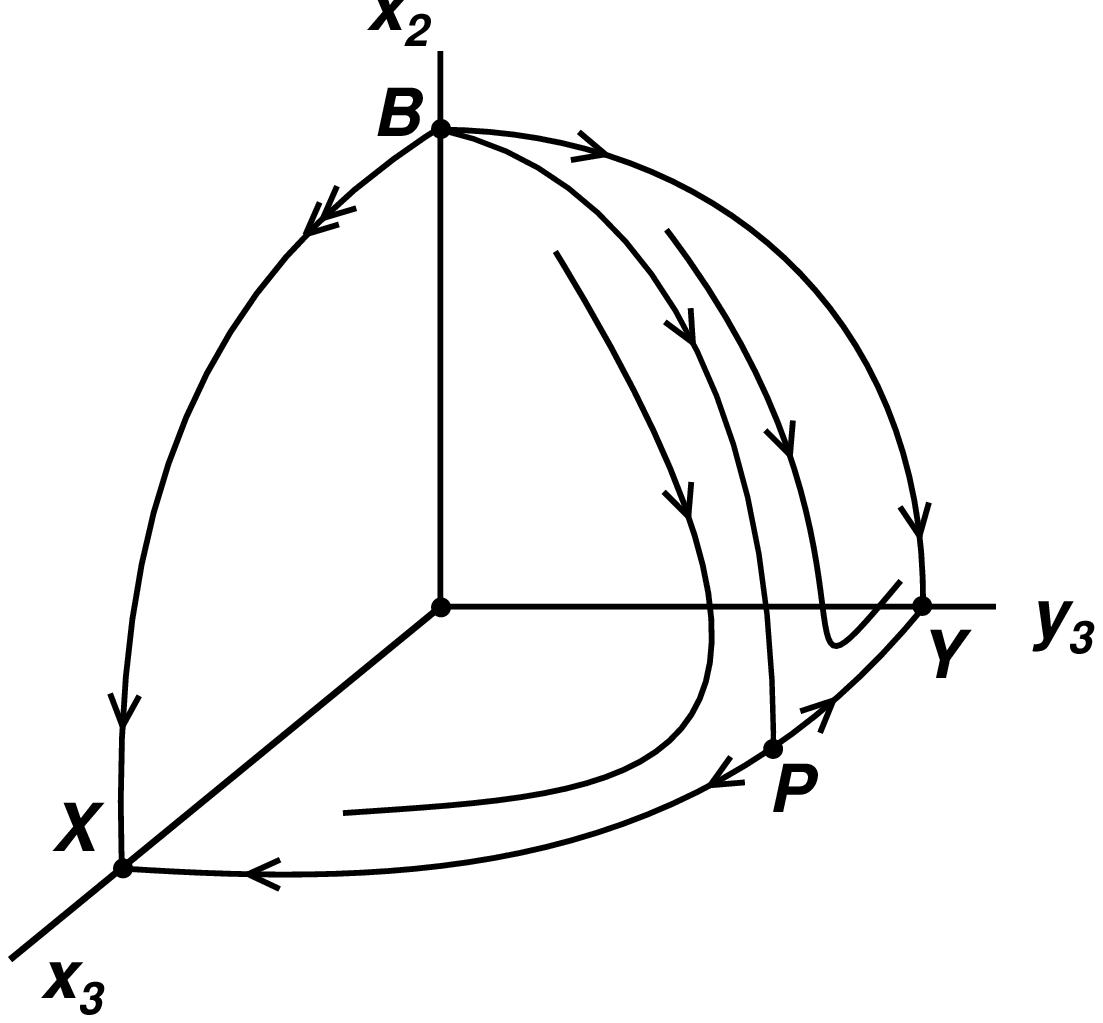,width=5cm}}\qquad
\subfigure[]{\epsfig{figure=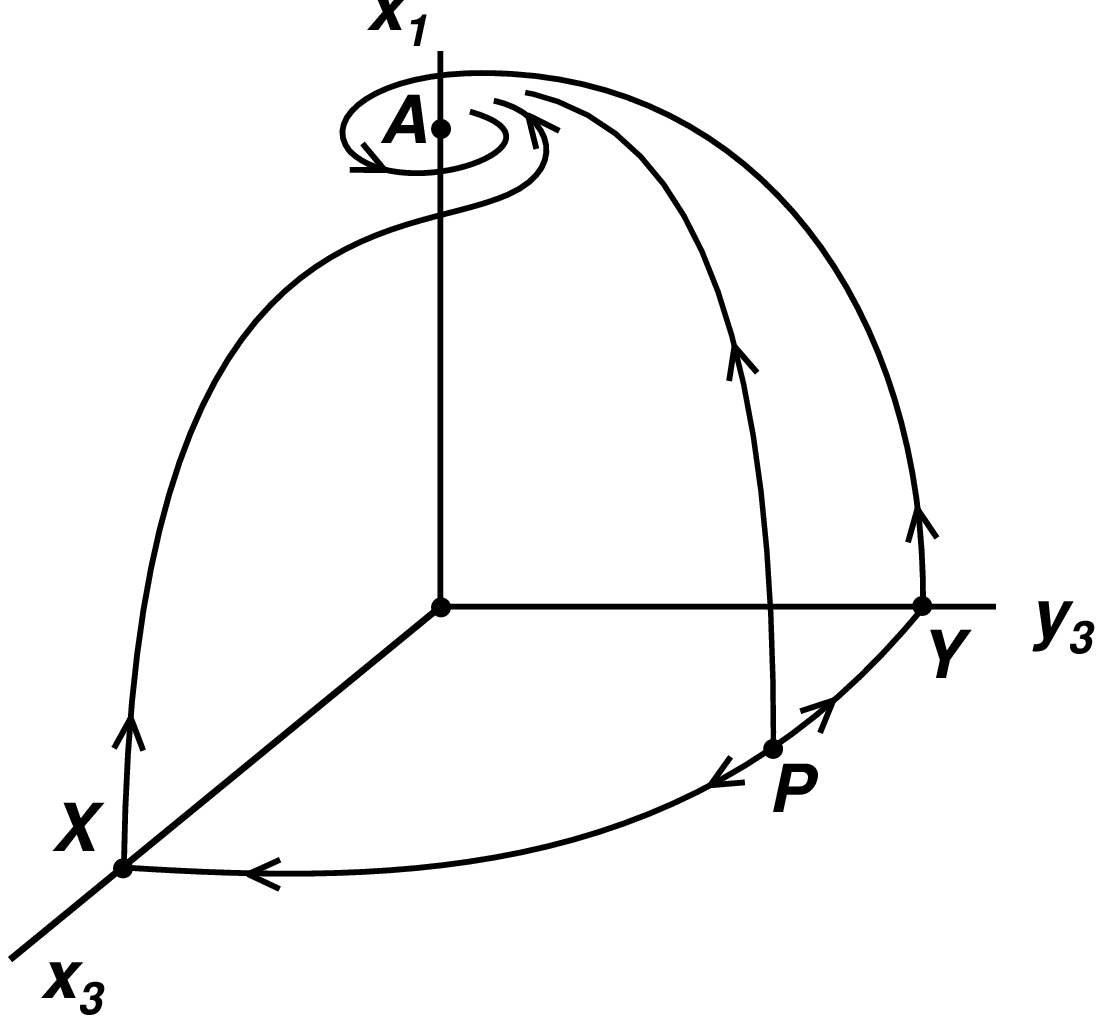,width=5cm}}
\end{center}
\caption{Dynamics within the subspaces invariant under the symmetries
$\kappa_1$, $\kappa_2$, $\kappa_3$ and their combinations. For clarity, only part of the
relevant subspaces are shown in panels (a), (c) and (d), with the dynamics in the
omitted parts being obtained by applying the symmetries.
 (a)~The invariant
plane $x_3=y_3=0$, showing the heteroclinic connection from $A$ to~$B$.
 (b)~The invariant plane $x_1=x_2=0$, showing the invariant circle~$C$ and the
equilibria $\pm{X}$, $\pm{Y}$, $\pm{P}$ and~$\pm{Q}$ that lie on~$C$.
 (c)~The subspace $x_1=0$ showing part of the two-dimensional unstable manifold
of~$B$ and part of the circle~$C$ in the $(x_3,y_3)$ plane. Most trajectories
leaving~$B$ go to either~$\pm{X}$ or~$\pm{Y}$, but some
isolated trajectories go to $\pm{P}$
or~$\pm{Q}$. For
convenience, the equilibria $\pm{X}$ and $\pm{Y}$ are chosen to lie on the
coordinate axes, with the eigenvectors of the corresponding linearised flow
at~$B$
aligned with the axes, but they are not constrained by symmetry to be there.
 (d)~The subspace $x_2=0$ showing spiralling of the unstable manifolds of $X$,
$Y$ and $P$ into~$A$. The unstable manifold of~$Q$ (not shown) behaves
similarly.
 In each subspace, the flow is strongly contracting in the radial direction.
}
 \label{fig:invsub}
 \end{figure}

We consider a system of ordinary differential equations,
$\dot{\mathbf{x}}={\mathbf{f}}(\mathbf{x})$, where
$\mathbf{x}=(x_1,x_2,x_3,y_3)\in\Rset^4$ and
${\mathbf{f}}:\Rset^4\rightarrow\Rset^4$ is a $\mathbf{C}^1$ vector-valued
function.  We assume this system is $\mathbb{Z}_2^3$-equivariant with the following
equivariance properties:
 \begin{equation}
 \kappa_i(\mathbf{f}(\mathbf{x})) =
 \mathbf{f}(\kappa_i(\mathbf{x})), \qquad i=1,2,3, \label{sym4p}
 \end{equation}
where
 \[ \begin{array}{rcl}
 \kappa_{1}:(x_{1},x_{2},x_{3},y_{3}) &\rightarrow
 &(-x_{1},x_{2},x_{3},y_{3}), \\
 \kappa_{2}:(x_{1},x_{2},x_{3},y_{3}) &\rightarrow
 &(x_{1},-x_{2},x_{3},y_{3}), \\
 \kappa_{3}:(x_{1},x_{2},x_{3},y_{3}) &\rightarrow &(x_{1},x_{2},-x_{3},-y_{3}) .
 \end{array}
 \]
These symmetries ensure the existence of dynamically invariant
subspaces in which robust saddle--sink
connections can occur. We make the following assumptions about the dynamics in
these subspaces (see Figure~\ref{fig:invsub}).

%%%%%%%%%%%%%
 %%%%%%%%%%%%%%

\begin{itemize}
\item{{\bf A1}:}
 There exist symmetry-related pairs of equilibria $\pm{A}$ and $\pm{B}$ on the
$x_1$ and $x_2$ coordinate axes, respectively. Within the invariant plane
$x_3=y_3=0$, $A$~is a saddle and $B$~is a sink and there is a heteroclinic
connection from $A$ to~$B$. See Figure~\ref{fig:invsub}(a).

\item{{\bf A2:}}
 There exist symmetry-related pairs of equilibria $\pm{X}$, $\pm{Y}$, $\pm{P}$
and $\pm{Q}$ in the invariant plane $x_1=x_2=0$.  Within this subspace,
$\pm{X}$ and $\pm{Y}$ are sinks, while $\pm{P}$ and $\pm{Q}$ are saddles. The
eight equilibria together with the heteroclinic connections between them make
up an invariant curve~$C$, which is topologically a circle. We hereafter refer to
$C$ as a circle, and we assume that $C$ can be parametrised by the
angle~$\theta_3$, the polar angle in the $(x_3,y_3)$-plane.
Note that the intersections of the stable manifolds of
$\pm{P}$ and $\pm{Q}$ with the invariant plane form the boundaries between the
basins of attraction of $\pm{X}$ and $\pm{Y}$ in the invariant plane. Only a small
part of each intersection is shown in Figure~\ref{fig:invsub}(b), to avoid giving a
misleading impression about the dynamics near the origin of the $(x_3,y_3)$-plane,
but each intersection curve in fact extends to the origin of the subspace.

\item{{\bf A3:}}
 Within the invariant subspace $x_1=0$, there exist two-dimensional manifolds
of saddle--sink connections from $B$ to $\pm{X}$ and $\pm{Y}$
(Figure~\ref{fig:invsub}(c)).  There are also one-dimensional (saddle--saddle or saddle--sink)
heteroclinic connections from $B$ to $\pm{P}$ and $\pm{Q}$ and from $\pm{P}$
and $\pm{Q}$ to $\pm{X}$ and $\pm{Y}$, as shown in Figure~\ref{fig:invsub}(c).
The unstable manifold of~$B$ is two-dimensional, and the stable manifolds of $\pm{X}$ and $\pm{Y}$
are each three-dimensional within the subspace.

\item{{\bf A4:}}
 Within the invariant subspace $x_2=0$, there exists a two-dimensional manifold
of saddle--sink connections from $\pm{X}$, $\pm{Y}$, $\pm{P}$ and $\pm{Q}$
to~$A$. Within this manifold, $A$~is a stable focus. A similar manifold connects the equilibria
on $C$ to $-A$. Apart from the
heteroclinic connections from $\pm{P}$ and $\pm{Q}$ to $\pm{X}$ and $\pm{Y}$,
the unstable manifolds of $\pm P$ and $\pm Q$ are contained in the stable manifolds of
$A$ and $-A$. There are no equilibria other than the origin and those mentioned above
lying in the subspace $x_2=0$. See Figure~\ref{fig:invsub}(d).

\item{{\bf A5:}}
 Equilibrium $B$ has real eigenvalues corresponding to dynamics in its
unstable manifold, and these eigenvalues are unequal.

\end{itemize}

Assumptions {\bf A1--A5} ensure the existence of the heteroclinic network shown
in Figure~\ref{fig:network}. The symmetries~$\kappa_1$ and $\kappa_2$ ensure
that $x_1$ and $x_2$ cannot change sign along a trajectory,
so we consider $x_1\geq0$ and
$x_2\geq0$ only. The complex eigenvalues at~$A$ enable both signs of $x_3$
and~$y_3$ to occur along trajectories. To simplify our analysis, we make the further
assumptions:

\begin{itemize}

\item{{\bf A6:}}
 At $A$ and~$B$, the directions of strongest
contraction lie along the coordinate axes $x_1$ and~$x_2$ respectively. At
each of $\pm{X}$, $\pm{Y}$, $\pm{P}$ and~$\pm{Q}$, the direction
of strongest contraction lies in the $(x_3,y_3)$ plane; this direction is
automatically transverse to the connections from $\pm{P}$ or $\pm{Q}$ to
$\pm{X}$ or~$\pm{Y}$.

\item{{\bf A7:}}
 The two expanding eigenvectors at~$B$ lie in the $x_3$ and $y_3$ directions.
Without loss of generality we assume that the eigenvalue in the $x_3$
direction is larger than that in the $y_3$ direction.
We also assume that the linearisation around~$A$, where there are complex
eigenvalues, is in Jordan form.

\item{{\bf A8:}}
 The equilibria $\pm{X}$ and $\pm{Y}$ are, respectively, on the $x_3$ and $y_3$
coordinate axes.

\end{itemize}

Note that we can always choose coordinates so that at least one of {\bf A7} and {\bf A8}
is satisfied,
but we assume both are satisfied in order to simplify the calculations. This has no
effect on our results.

Thus the overall network is $A\rightarrow B\rightarrow C\rightarrow A$, where,
within~$C$, trajectories can visit any of $\pm X$, $\pm Y$, $\pm
P$ and $\pm Q$, although only in certain orders as indicated in
Figures~\ref{fig:network} and \ref{fig:invsub}. All
cycles in the network contain either three or four equilibria.

 {\bf Definition.}
For a trajectory $\phi(t)$ close to the network, we define the {\it itinerary}
of $\phi(t)$ to be the sequence $\{\xi_j\}$ of equilibria visited. That is, $\{\xi_j\}$
is the itinerary of the trajectory $\phi(t)$ if
there exists an increasing sequence of times $\{t_j\}$ such that the distance from
$\phi(t_j)$ to the equilibrium $\xi_j$ is less than some small constant.
For a trajectory that stays close to a single heteroclinic cycle, the itinerary will be
a periodic sequence, with the (minimal) length of the repeating segment of the sequence being equal to the
number of equilibria in the cycle.

In this paper we use the following definition for switching, defined for a particular trajectory close to a network. Note that Aguiar {\it et al.}~\cite{Aguiar2005}
define switching as a property of a network, not individual trajectories.

{\bf Definition.}
We say a trajectory {\em switches} if, as
$t\to\infty$, the itinerary is not eventually a periodic sequence with minimal
period three or four, that is, the trajectory does not eventually remain near a single
cycle of the network. In this definition, we distinguish between conjugate
equilibria, that is, $X$ and $-X$ and so on.

%%%%%%%%%%%%%%%%%%%%%%%%%%%%%%
\subsection{Coordinates, cross-sections, and local maps}
\label{sec:coordcross}
%%%%%%%%%%%%%%%%%%%%%%%%%%%%%%%

In this section, we define the coordinates, cross-sections, and local
maps required for modelling
the dynamics near our heteroclinic network.

Near $A$ and~$B$, we define local coordinates that place
the equilibrium at the origin. Assumption~{\bf A7} guarantees that
the coordinate axes are aligned with the
eigenvectors of the relevant linearised system.
We use polar coordinates when it is more convenient:
$(x_3,y_3)$ becomes $(r_3,\theta_3)$, where $x_3=r_3\cos\theta_3$ and
$y_3=r_3\sin\theta_3$. We write $x_i$
or~$y_i$ if the local coordinate is the same as the corresponding global
coordinate. At $A$ and $B$ we use $u_1$ and $u_2$ for the radial
coordinate relative to the equilibrium points.
At the invariant circle~$C$,
we use a $\theta_3$-dependent coordinate transformation to define a
radial coordinate~$u_3$.

Near $A$, the linearised flow is given by:
 \begin{equation}
 \dot{u}_1=-\rA u_1,\
 \dot{x}_2=\eA x_2,\
 \dot{x}_3=-\cA x_3 - \omega y_3,\
 \dot{y}_3=\omega x_3 - \cA y_3,
 \label{eq:linA}
 \end{equation}
 where $\rA$, $\eA$, $\cA$ and $\omega$ are positive constants.
In polar coordinates, the $\dot{x}_3$ and $\dot{y}_3$ equations give
$\dot{r}_3=-\cA r_3$ and $\dot{\theta}_3=\omega$.

Cross-sections near~$A$ are defined as:
 \begin{equation}
 \label{HA}
 \begin{array}{lcl}
 \HAin
 &\equiv &
 \{(u_1,x_2,r_3,\theta_3)\, \big|\,  |u_1|<h, 0\leq x_2<h,
 r_3=h, 0\leq\theta_3<2\pi \}, \\
 \HAout
 &\equiv &
 \{(u_1,x_2,r_3,\theta_3)\, \big|\, |u_1|<h, x_2=h,
 0\leq r_3<h, 0\leq\theta_3<2\pi\}.
 \end{array}
 \end{equation}
Here~$0<h\ll 1$ is a parameter small enough that the cross-sections lie within
the region of approximate linear flow near~$A$ (and similarly near $B$ and $C$, as
required below).

The flow near $A$ induces a map $\phi_A: \HAin \to \HAout$, which is obtained to
lowest order by integrating equations~(\ref{eq:linA}):
 \begin{equation}
 \label{phiA}
 \phi_{A}(u_1,x_2,h,\theta_3) =
 \left(u_1\left(\frac{x_2}{h}\right)^{\frac{\rA}{\eA}},h,
       h\left(\frac{x_2}{h}\right)^{\deltaA},
       \theta_3-\frac{\omega}{\eA}\log\left(\frac{x_2}{h}\right)
 \right)
 \end{equation}
where $\deltaA=\frac{\cA}{\eA}$.

Near $B$, the linearised flow is:
 \begin{equation}
 \dot{x}_1=-\cB x_1,\
 \dot{u}_2=-\rB u_2,\
 \dot{x}_3= \eBx x_3,\
 \dot{y}_3= \eBy y_3,
 \label{eq:linB}
 \end{equation}
 where $\rB$, $\eBx$, $\eBy$, $\cB$ are positive constants.
From~{\bf A7}, we have $\eBx>\eBy$.

Cross-sections near~$B$ are defined as:
 \begin{equation}
 \label{HB}
 \begin{array}{lcl}
 \HBin
 &\equiv &
 \{(x_1,u_2,r_3,\theta_3)\, \big|\,  x_1=h, |u_2|<h,
 0\leq r_3<h, 0\leq\theta_3<2\pi \}, \\
 \HBout
 &\equiv &
 \{(x_1,u_2,r_3,\theta_3)\, \big|\, 0\leq x_1<h, |u_2|<h,
 r_3=h, 0\leq\theta_3<2\pi\}.
 \end{array}
 \end{equation}

The flow induces a map $\phi_B: \HBin \to \HBout$, which is obtained to lowest
order by integrating equations~(\ref{eq:linB}). The map cannot be written down
explicitly, but is computed as follows. First,
the $\dot{x}_3$ and $\dot{y}_3$ equations are solved:
 $$
 x_3(t)=r_3(0)\cos\theta_3(0)\, e^{\eBx t},
 \qquad
 y_3(t)=r_3(0)\sin\theta_3(0)\, e^{\eBy t},
 $$
 where $r_3(0)$ and $\theta_3(0)$ are the initial values of the radial
 coordinates (i.e., on $\HBin$).
The trajectory crosses $\HBout$ when $r_3(t)=h$, so the
transit time~$T_B$ is found by solving the equation
 \begin{equation}
 \label{eq:TB}
 \left(\frac{h}{r_3(0)}\right)^2 = \cos^2\theta_3(0) \, e^{2\eBx T_B} +
                                \sin^2\theta_3(0) \, e^{2\eBy T_B}
 \end{equation}
for~$T_B$ in terms of $r_3(0)$ and $\theta_3(0)$. This yields
the local map
$\phi_B:\HBin\rightarrow\HBout$:
 \begin{equation}
 \label{phiB}
 \phi_B(h,u_2,r_3,\theta_3) =
 \left(h e^{-\cB T_B},
       u_2 e^{-\rB T_B},
       h,
       \tan^{-1}\left(\tan(\theta_3) e^{(\eBy-\eBx)T_B}\right)
       \right).
 \end{equation}
For later convenience, we define
$\deltaBx\equiv\deltaBmin=\frac{\cB}{\eBx}$ and
$\deltaBy\equiv\deltaBmax=\frac{\cB}{\eBy}$.

The treatment of the dynamics near the invariant circle~$C$ is more complicated.
We assumed in {\bf A2} that
$C$ can be parameterised by the angle~$\theta_3$.
The rate of relaxation
onto~$C$ is controlled by the $\theta_3$-dependent
eigenvalue $-\rC(\theta_3)$.
The assumption of strong contraction in the radial ($r_3$) direction
({\bf A6}) means that the
dynamics on~$C$ of $\theta_3$ can be described by a one-dimensional nonlinear ODE of the
form $\dot{\theta}_3=g(\theta_3)$. The presence of
 $\pm{X}$ and $\pm{Y}$ on the coordinate axes will require
$g(0)=g(\pi/2)=g(\pi)=g(3\pi/2)=0$. The presence of  $\pm{P}$ and
$\pm{Q}$ will require further zeroes of~$g$.
This results in the flow near~$C$ being given by:
 \begin{equation}
 \dot{x}_1= \eC(\theta_3) x_1,\
 \dot{x}_2=-\cC(\theta_3) x_2,\
 \dot{u}_3=-\rC(\theta_3) u_3,\
 \dot{\theta}_3= g(\theta_3),
 \label{eq:linC}
 \end{equation}
 where $\rC$, $\eC$ and $\cC$ are positive functions of~$\theta_3$.

Cross-sections near~$C$ are defined as:
 \begin{equation}
 \label{HC}
 \begin{array}{lcl}
 \HCin
 &\equiv &
 \{(x_1,x_2,u_3,\theta_3)\, \big|\,  0\leq x_1<h, x_2=h,
 |u_3|<h, 0\leq\theta_3<2\pi \}, \\
 \HCout
 &\equiv &
 \{(x_1,x_2,u_3,\theta_3)\, \big|\, x_1=h, 0\leq x_2<h,
 |u_3|<h, 0\leq\theta_3<2\pi\}.
 \end{array}
 \end{equation}
There is a continuum of heteroclinic connections from $B$ to the various
equilibrium points in~$C$, and defining the cross-sections in this way allows
us to keep track of all these connections.

The flow induces a map  $\phi_C: \HCin \to \HCout$. As in the case of the flow
past~$B$, we cannot write down the map explicitly, but it is computed as
follows. First, the $\dot{\theta}_3$ equation is solved using an
initial condition~$\theta_3(0)$, yielding $\theta_3(t)$.
Then the $\dot{x}_1$ and $\dot{x}_2$ equations are solved:
 $$
 x_1(t)=x_1(0) \exp\left(\int_0^t\eC(\theta_3(t'))\,dt'\right),
 \qquad
 x_2(t)= h  \exp\left(-\int_0^t\cC(\theta_3(t'))\,dt'\right).
 $$
The trajectory crosses $\HCout$ when $x_1(t)=h$, so the
transit time~$T_C$ can be found in principle by solving
 \begin{equation}
 \label{eq:TC}
 \int_0^{T_C}\eC(\theta_3(t'))\,dt' = -\log\left(\frac{x_1(0)}{h}\right)
 \end{equation}
for~$T_C$ in terms of the initial values $x_1(0)$ and~$\theta_3(0)$ on~$\HCin$.
Then the local map $\phi_C:\HCin\rightarrow\HCout$ is given by
 \begin{equation}
 \label{phiC}
 \phi_C(x_1,h,u_3,\theta_3) =
 \left(h,  h \exp\left(-\int_0^{T_C}\cC(\theta_3(t'))\,dt'\right),
           u_3(T_C),
          \theta_3(T_C)
       \right),
 \end{equation}
where $u_3(T_C)=u_3 \exp\left(-\int_0^{T_C}\rC(\theta_3(t'))\,dt'\right)$.
For later convenience, we define $\deltaCX$, $\deltaCY$, $\deltaCP$
and~$\deltaCQ$,  to be the ratio $\frac{\cC(\theta_3)}{\eC(\theta_3)}$
evaluated at the points~$X$, $Y$, $P$ and~$Q$, respectively.

Neither of the local maps~$\phi_B$ and $\phi_C$ can be written down explicitly.
In the case of~$\phi_B$, the obstruction is only
that we cannot write down an explicit solution of (\ref{eq:TB}) for the transit
time~$T_B$. In the case of~$\phi_C$, the nonlinear evolution of $\theta_3$
within~$C$ cannot be written down explicitly. However, in both cases, we will
be able to give bounds on some properties of the trajectories, and this turns out
to be sufficient for the purposes of determining stability and switching properties
of the network.

%%%%%%%%%%%%
\subsection{Global maps}
\label{sec:global}
%%%%%%%%%%%%%

We construct global maps $\Psi_{ij}$ to approximate the dynamics near the
heteroclinic connections between $A$, $B$ and $C$. In
each case, we linearise the dynamics about the unstable manifold leaving the
invariant set, taking into account the fact that the unstable manifold of~$A$
is one-dimensional, but the unstable manifolds of $B$ and $C$
are two-dimensional.
We make use of the equivariance of the vector field in our map construction.

The simplest of the global maps is $\Psi_{AB}:\HAout\to\HBin$. The heteroclinic
connection from $A$ to $B$ intersects $\HAout$ at
$(u_1,x_2,x_3,y_3)=(0,h,0,0)$, and
intersects $\HBin$ at $(x_1,u_2,x_3,y_3)=(h,\epsilon_B,0,0)$, for a small
constant~$\epsilon_B$. Generically, $\epsilon_B\ne 0$ and we assume that
this is the case in the following. Near the heteroclinic connection the map expressed
in cartesian coordinates is
at lowest order an affine linear transformation. Converting to polar coordinates, this yields,
at leading order:
 \begin{equation}
 \label{PsiAB}
 \Psi_{AB}(u_1,h,r_3,\theta_3) = (h,\epsilon_B,
 D_B(\theta_3)r_3, {\bar\theta}_B(\theta_3)),
 \end{equation}
where $D_B(\theta_3)$ is an order-one function of~$\theta_3$
that indicates how the small variable~$r_3$ is scaled in the transition from $A$ to~$B$,
and ${\bar\theta}_B(\theta_3)$ is an order-one function of~$\theta_3$.
The invariance of this map under the symmetry $\kappa_3$
ensures that there is no constant term in the $r_3$ component.
The overall effect of this map is to multiply~$r_3$ by an order-one
amount~$D_B$, and to rigidly rotate the angle~$\theta_3$.

The unstable manifold of~$B$ is
two-dimensional; it intersects $\HBout$ at
$(x_1,u_2,r_3,\theta_3)=(0,0,h,\theta_3)$, for
$0\leq\theta_3<2\pi$, and it intersects $\HCin$ at
$(x_1,x_2,u_3,\theta_3)=(0,h,\epsilon_C(\theta_3),{\bar\theta}_C(\theta_3))$,
where $\epsilon_C$ is a
small function of~$\theta_3$ and ${\bar\theta}_C$ is an order-one function
of~$\theta_3$. For small~$x_1$ and~$u_2$, we have at leading order:
 \begin{equation}
 \label{PsiBC}
 \Psi_{BC}(x_1,u_2,h,\theta_3) =
   \left(D_C(\theta_3)x_1, h,
    \epsilon_C(\theta_3),
    {\bar\theta}_C(\theta_3)\right),
 \end{equation}
where $D_C(\theta_3)$ is an order-one function of~$\theta_3$.
Here $\epsilon_C(\theta_3)$ plays a similar role to the constant
$\epsilon_B$ in (\ref{PsiAB}), except that it takes on a different constant
value for each heteroclinic connection and so is a function of $\theta_3$.
As with $\epsilon_B$, $\epsilon_C(\theta_3)$ is generically non-zero
and we assume that $\epsilon_C(\theta_3) \ne 0$ for any $\theta_3$.

The
effect of (\ref{PsiBC}) is to multiply the small variable~$x_1$ by an
order-one function of~$\theta_3$, and to map the outgoing angle $\theta_3$ to
an incoming angle~$\bar{\theta}_C$. Unlike in the case of $\Psi_{AB}$, the
effect of $\bar{\theta}_C$ need not be a rotation.

For the global map $\Psi_{CA}:\HCout\to\HAin$, we also use
$(r_3,\theta_3)$ rather than~$(x_3,y_3)$. The unstable manifold of $C$ is
two-dimensional; it intersects $\HCout$ at $(x_1,x_2,u_3,\theta_3)=(h,0,0,\theta_3)$, where
$0\leq\theta_3<2\pi$, and it intersects $\HAin$ at
$(u_1,x_2,r_3,\theta_3)=(\epsilon_A(\theta_3),0,h,{\bar\theta}_A(\theta_3))$, where $\epsilon_A$ is a
small function of~$\theta_3$, and ${\bar\theta}_A$ is an order-one function
of~$\theta_3$. For small~$x_2$ and~$u_3$, we have:
 \begin{equation}
 \label{PsiCA}
 \Psi_{CA}(h,x_2,u_3,\theta_3) =
   \left(\epsilon_A(\theta_3),
         D_A(\theta_3)x_2, h,
    {\bar\theta}_A(\theta_3)\right),
 \end{equation}
where $D_A(\theta_3)$ is an order-one function of~$\theta_3$.
The
effect of this map is to multiply the small variable~$x_2$ by an order-one
function of~$\theta_3$, and to map the outgoing angle $\theta_3$ to an incoming
angle~$\bar{\theta}_A$.
As in the case of $\Psi_{BC}$, the effect of
$\bar{\theta}_A$ need not be a rotation.

\section{Analysis of the maps}
\label{sec:analysis}

By composing the local and global maps in an appropriate order, we construct
return maps that approximate the dynamics near the cycles in our network. We
are interested in finding conditions under which the network as a whole is
attracting, and in describing the switching properties of trajectories as they
travel around the network.
We are particularly interested in  trajectories that
repeatedly visit both ${X}$ (or~$-X$) and~${Y}$ (or~$-Y$).

In our analysis, as in~\cite{Kirk2008},
we make use of the observation that at each cross-section, the four
variables play distinct roles. Two variables are unimportant: the one
that is equal to~$h$, and the one in the radial direction. Of the other two
variables, one is \emph{small} and measures the distance from an invariant
subspace, and the other is the order-one \emph{angle}~$\theta_3$. At each
cross-section, the roles change, but there are always small and angle variables.

 %%%%%%%%%%%%%%%%
\subsection{Stability results}
\label{sec:stability}

To show that the network as a whole is attracting, we must find
conditions under which the small variable decreases each time around the
network. In order to do this, we bound this variable over all possible values
of the angle variable. This means that we need to take into account the details
of which part of $C$ is visited by the trajectory. We are unable to compute
the stability result by direct computation of a return map, since the local
maps $\phi_B$ and $\phi_C$ are only known implicitly, but the lengthy
computation below achieves the same result.

We start on $\HAin$ at $(u_1,x_2,h,\theta_3)$, and consider the effects of
maps $\phi_A$, $\Psi_{AB}$, $\phi_B$, $\Psi_{BC}$, $\phi_C$ and $\Psi_{CA}$ in
turn. We assume that $x_2\ll1$. After~$\phi_A$ and $\Psi_{AB}$, we arrive on
$\HBin$ at:
 $$
 \left(
 h, \epsilon_B,
 hD_B\left(\theta_3-\frac{\omega}{\eA}\log\left(\frac{x_2}{h}\right)\right)\left(\frac{x_2}{h}\right)^{\deltaA},
 \bar{\theta}_B\left(\theta_3-\frac{\omega}{\eA}\log\left(\frac{x_2}{h}\right)\right)
 \right),
 $$
where we have discarded all higher-order corrections. For
convenience, we label the values of $r_3$ and $\theta_3$ on $\HBin$
as $\riiiBin$ and
$\thetaBin$, with a similar convention on other cross-sections. We define $\DBmax$ and $\DBmin$ to be the maximum and
minimum values of $D_B$ taken over all values of~$\theta_3$. Then we
can bound the small variable~$\riiiBin$ by
 \begin{equation}
 \label{ineq:r3B}
 h\DBmin\left(\frac{x_2}{h}\right)^{\deltaA}
 \leq
 \riiiBin
 \leq
 h\DBmax\left(\frac{x_2}{h}\right)^{\deltaA}.
 \end{equation}

Next, we consider the effect of maps~$\phi_B$ and~$\Psi_{BC}$. Recall from ({\bf
A7}) that
$\eBx>\eBy$. From~(\ref{eq:TB})
we can bound $T_B$:
 $$
 -\frac{1}{\eBx}\log\left(\frac{\riiiBin}{h}\right)
 \leq
 T_B
 \leq
 -\frac{1}{\eBy}\log\left(\frac{\riiiBin}{h}\right).
 $$
As a result, the small variable~$\xiBout$ on $\HBout$ is bounded by
 $$
 h\left(\frac{\riiiBin}{h}\right)^{\deltaBy}
 \leq \xiBout \leq
 h\left(\frac{\riiiBin}{h}\right)^{\deltaBx}.
 $$
Note that $\deltaBx<\deltaBy$.
After~$\Psi_{BC}$, trajectories enter~$\HCin$ at:
 $$
 \left(
 D_C\left(\thetaBout\right)\xiBout, h,
 \epsilon_C(\theta_3),
 \bar{\theta}_C\left(\thetaBout\right)
 \right),
 $$
where we have discarded all higher-order corrections.
We define $\DCmax$ and $\DCmin$ to be the maximum and
minimum values of $D_C$ taken over all values of $\theta_3$. Then we
can bound~$\xiCin$ by
 \begin{equation}
 \label{ineq:x1C}
 h\DCmin\left(\frac{\riiiBin}{h}\right)^{\deltaBy}
 \leq
 \xiCin
 \leq
 h\DCmax\left(\frac{\riiiBin}{h}\right)^{\deltaBx}.
 \end{equation}

Finally, we consider the effects of maps~$\phi_C$ and $\Psi_{CA}$. The first of
these is the most complicated as trajectories can enter the neighbourhood
of~$C$ close to any of the equilibrium points $\pm{X}$, $\pm{Y}$, $\pm{P}$
or~$\pm{Q}$ (or in between), and can similarly exit the neighbourhood of~$C$
close to any of the equilibrium points $\pm{X}$, $\pm{Y}$, $\pm{P}$ or~$\pm{Q}$
(or in between). We take all possibilities into account and derive a bound on
the exit value of the small variable~$x_2$.

\begin{figure}
\psfrag{X}{$X$}
\psfrag{Y}{$Y$}
\psfrag{C}{$C$}
\psfrag{P}{$P$}
\psfrag{a}{(a)}
\psfrag{b}{(b)}
\psfrag{c1}{(c)}
\psfrag{Hin}{$\HCin$}
\psfrag{Hout}{$\HCout$}
\psfrag{tx}{$\theta_3^X$}
\psfrag{txh}{$\theta_3^X+h$}
\psfrag{tph}{$\theta_3^P-h$}
\psfrag{tp}{$\theta_3^P$}
\begin{center}
\epsfig{figure=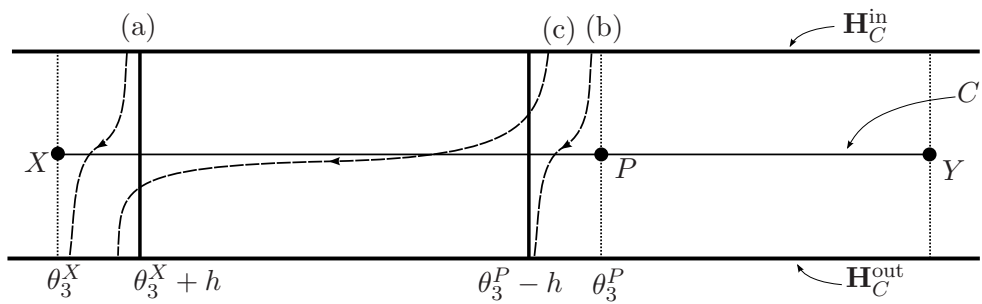,width=\textwidth}
\end{center}
\caption{\label{fig:Ctraj}Schematic showing three main possibilities of how
trajectories pass through the region near $C$. Trajectories are shown as dashed
lines, Poincar\'e sections as solid bold lines. Equilibria are indicated by
dots. The labels (a), (b) and (c) identify trajectories representative of three
cases discussed in the text.}
 \end{figure}

To simplify the discussion, we consider in detail only the case of trajectories
arriving at~$C$ between $X$ and~$P$; the other cases can easily be deduced from
this one. Within $C$, $X$ is stable, with a stable
eigenvalue~$-\lambda_X$, and $P$ is unstable, with an unstable
eigenvalue~$\lambda_P$, with $\lambda_X,\lambda_P>0$. The corresponding
eigenvectors are within the $(x_3,y_3)$ plane, and are tangent to~$C$ at~$X$
and~$P$. To aid the analysis, we consider cross-sections orthogonal to~$C$, at
$\theta_3=\theta_3^X+h$ and $\theta_3=\theta_3^P-h$, where $\theta_3^X$ and
$\theta_3^P$ are the $\theta_3$~coordinates of $X$ and $P$ respectively, and
$h$ is as before (see Figure~\ref{fig:Ctraj}). Note that we have defined our coordinates
so that $\theta_3^X=0$, but for clarity of the following discussion we leave this constant in symbolic form.

%%%%%%%%%%%%%
 %%%%%%%%%%%%%%

There are three main possibilities, indicated in Figure~\ref{fig:Ctraj}:
(a)~the trajectory crosses~$\HCin$ near~$X$
and must therefore cross~$\HCout$ near~$X$ as well; (b)~the trajectory
crosses~$\HCin$ near~$P$ and also crosses~$\HCout$ near~$P$; and (c)~the
trajectory crosses~$\HCin$ near~$P$ and leaves the neighbourhood of~$P$,
heading towards~$X$, and so crosses~$\HCout$ near~$X$. There are the additional
possibilities that the trajectory crosses $\HCin$ or $\HCout$ in between $X$
and~$P$; we discuss these cases below. Throughout this discussion, we disregard
the radial coordinate~$r_3$.

Case~(a) is straightforward: the flow near~$X$ is given by the linearisation
of~(\ref{eq:linC}) around~$X$:
 \begin{equation}
 \label{eq:linCX}
 \dot{x}_1 = \eC(\theta_3^X)x_1,\
 \dot{x}_2 = -\cC(\theta_3^X)x_2,\
 \dot{\theta}_3 = -\lambda_X(\theta_3-\theta_3^X),
 \end{equation}
and so the linearised map near $X$ takes an incoming point
$(\xiCin, h, \thetaCin)$ to
 $$
(x_1,\xiiCout,\thetaCout)= \left(h,h\left(\frac{\xiCin}{h}\right)^{\deltaCX},
 \theta_3^X + (\thetaCin-\theta_3^X)\left(\frac{\xiCin}{h}\right)^{\frac{\lambda_X}{\eC(\theta_3^X)}}
 \right),
 $$
where $\deltaCX$ is the ratio of contracting to expanding eigenvalues
evaluated at~$X$.

Similarly, case~(b) is treated by linearising (\ref{eq:linC}) around~$P$, and
results in exit values
$$
(x_1,\xiiCout,\thetaCout)=
 \left(h,h\left(\frac{\xiCin}{h}\right)^{\deltaCP},
 \theta_3^P + (\thetaCin-\theta_3^P)\left(\frac{\xiCin}{h}\right)^{\frac{-\lambda_P}{\eC(\theta_3^P)}}
 \right),
 $$
where $\deltaCP=\frac{c_{C}(\theta_3^P)}{e_{C}(\theta_3^P)}$ is the ratio of contracting to expanding eigenvalues
evaluated at~$P$. The condition that the trajectory crosses $\HCout$ before it
crosses $\theta_3=\theta_3^P-h$ amounts to:
 $$
 \left|\thetaCin-\theta_3^P\right|
 \left(\frac{\xiCin}{h}\right)^{\frac{-\lambda_P}{\eC(\theta_3^P)}} < h,
 $$
that is, the trajectory must enter~$C$ close enough to~$P$ that the small
variable~$x_1$  grows to size~$h$ before the angular separation
$\theta_3-\theta_3^P$ grows to size~$h$.

Finally, in case~(c), there are three stages: linearised dynamics near~$P$, a
jump from $P$ to $X$, and linearised dynamics near~$X$. The time for the first
stage is found by setting $\theta_3(t)=\theta_3^P-h$, and then the  $(x_1,x_2)$ coordinates on
this section are:
 $$
 \left(
 \xiCin\left|\frac{\thetaCin-\theta_3^P}{h}\right|^{-\frac{\eC(\theta_3^P)}{\lambda_P}},
 h        \left|\frac{\thetaCin-\theta_3^P}{h}\right|^{ \frac{\cC(\theta_3^P)}{\lambda_P}}
 \right).
 $$
Then there is a jump from $\theta_3=\theta_3^P-h$ to
$\theta_3=\theta_3^X+h$, during which $x_1$ and $x_2$ change by a factor
of $\DXone$ and $\DXtwo$ respectively. The values of $\DXone$ and
$\DXtwo$ depend on~$h$. Lastly, there is the linearised
dynamics near~$X$, which results in an exit value of $x_2$ given by:
 \begin{equation}
 \label{eq:exitx2}
 \xiiCout= h \DXtwo \left(\frac{\DXone\xiCin}{h}\right)^{\deltaCX}
 \left|\frac{\thetaCin-\theta_3^P}{h}\right|
      ^{\frac{\eC(\theta_3^P)}{\lambda_P}\left(\deltaCP-\deltaCX\right)}.
 \end{equation}
This case only occurs if
 \begin{equation}\label{eq:PXcond}
 \left|\frac{\thetaCin-\theta_3^P}{h}\right|
      ^{\frac{\eC(\theta_3^P)}{\lambda_P}}
 > \frac{\xiCin}{h}
 \end{equation}
(otherwise we would be in case~(b)). This condition allows us to bound the
exit values of $x_2$.
Note first that $\left|\frac{\thetaCin-\theta_3^P}{h}\right| ^{\frac{\eC(\theta_3^P)}{\lambda_P}}<1$.
Then if $\deltaCP>\deltaCX$, from~(\ref{eq:exitx2}) we have that $\xiiCout$ is bounded above by a constant times
$\left(\xiCin\right)^{\deltaCX}$. Additionally, using~\eqref{eq:PXcond} in~(\ref{eq:exitx2}), we have that $\xiiCout$ is bounded below by a constant times $\left(\xiCin\right)^{\deltaCP}$. Similar considerations in the case that $\deltaCP<\deltaCX$ give additional constraints which altogether result in:
 $$
 h\DXmin\left(\frac{\xiCin}{h}\right)^
  {\max\left(\deltaCP,\deltaCX\right)}
 \leq
 \xiiCout
 \leq
 h\DXmax\left(\frac{\xiCin}{h}\right)^
  {\min\left(\deltaCP,\deltaCX\right)},
 $$
where $\DXmin$ and $\DXmax$ are constants that depend on $\DXone$
and $\DXtwo$ and some exponents.

The same analysis can be used in the cases where trajectories enter or leave
the neighbourhood of~$C$ in between the neighbourhoods of~$P$ and~$X$, with
only minor alterations of the values of the constants $\DXmin$
and~$\DXmax$. Then all possibilities (a), (b) and~(c) can be assembled,
as well as including trajectories that visit the equilibria $Y$ and $Q$ as
well, and the map~$\Psi_{CA}$ can be applied. All this results in a bound on the
value of $\xiiAin$ at $\HAin$ of the form:
 \begin{equation}
 \label{ineq:x2A}
 h\DAmin\left(\frac{\xiCin}{h}\right)^{\deltaCmax}
 \leq
 \xiiAin
 \leq
 h\DAmax\left(\frac{\xiCin}{h}\right)^{\deltaCmin},
 \end{equation}
where we interpret $\deltaCmax$ as
$\max\left(\deltaCX,\deltaCY,\deltaCP,\deltaCQ\right)$, and
similarly $\deltaCmin$, and the constants $\DAmin$ and $\DAmax$ are the
smallest and largest of all the constants in the local and global parts of the
maps.

Recall that the trajectory started on $\HAin$ with particular values of $x_2$
and $\theta_3$. In computing these bounds, the value of $\theta_3$ has been
lost, but inequalities (\ref{ineq:r3B}), (\ref{ineq:x1C}) and (\ref{ineq:x2A})
provide the smallest and largest possible values of $x_2$ once the trajectory
returns to~$\HAin$\/:
 \begin{equation}
 \label{ineq:x2all}
 h\DABCmin\left(\frac{x_2}{h}\right)^{\deltamax}
 \leq
 \xiiAin
 \leq
 h\DABCmax\left(\frac{x_2}{h}\right)^{\deltamin}.
 \end{equation}
Here $\deltamax=\deltaA\deltaBmax\deltaCmax$,
      $\deltamin=\deltaA\deltaBmin\deltaCmin$, and all the
constants have been amalgamated into $\DABCmin$ and~$\DABCmax$.

We have thus established a condition for asymptotic stability or instability of
the network.
If $\deltamin>1$, then a trajectory starting close enough to the network will
return closer to the network (with a smaller value of~$x_2$) regardless of
which itinerary it takes and regardless of the values of $\DABCmin$
and~$\DABCmax$, and so the network is asymptotically stable. If $\deltamax<1$,
then a trajectory starting close to the network will return further away from
the network (with a larger value of~$x_2$) regardless of which itinerary it
takes, and so the network is unstable. The values of the constants can be
scaled away in both these cases.

These conditions for asymptotically stability and instability are as expected:
the product of the ratio of the contracting to expanding eigenvalues should be
greater or less than one regardless of the itinerary. The more interesting and
complicated case is when $\deltamin<1<\deltamax$, in which case it appears that whether
there is net contraction or expansion depends on the itinerary. We present some
numerical results relevant to this case in section~\ref{sec:numeg}. These
results suggest that the network may be essentially asymptotically stable
or unstable, depending on which of the routes through the network is
responsible for $\deltamax>1$ and which is responsible for~$\deltamin<1$.

\subsection{Switching near the network}
\label{sec:switching}

In this section, we show that close enough to the network, there are
trajectories that, over the course of two circuits around the network, visit any
combination of the equilibrium points within~$C$ in any order.
This occurs whether or not the
network is asymptotically stable. We also show that when the
network is asymptotically stable, most trajectories repeatedly visit both~$X$
and $-X$ as they approach the network. On the assumption that the
complex eigenvalues at~$A$ mix trajectories effectively, we estimate how often
trajectories visit~$\pm{Y}$ and show that, when the network is
asymptotically stable, visits to~$\pm{Y}$ become rare as trajectories approach
the network. Finally, we show that interesting switching dynamics might
be possible in the case $\deltamin<1<\deltamax$, when some parts of the network
are attracting and other parts are repelling.

%%%%%%%%%%%%%%%%%%%
\begin{figure}
\psfrag{X}{$X$}
\psfrag{mX}{$-X$}
\psfrag{Y}{$Y$}
\psfrag{PX}{$P\rightarrow X$}
\psfrag{P}{$P$}
\psfrag{PY}{$P\rightarrow Y$}
\psfrag{Q}{$Q$}
\psfrag{QY}{$Q\rightarrow Y$}
\psfrag{QX}{$Q\rightarrow -X$}
\psfrag{x3}{$x_3$}
\psfrag{y3}{$y_3$}
\psfrag{r3}{\small{$r_3=h$}}
\psfrag{r31}{\small{$r_3=10^{-1}h$}}
\psfrag{r32}{\small{$r_3=10^{-2}h$}}
\psfrag{t0}{$\theta_3=0$}
\psfrag{tp2}{$\theta_3=\frac{\pi}{2}$}
\psfrag{tp}{$\theta_3=\pi$}
\begin{center}
\subfigure[]{\includegraphics[width=8.7cm]{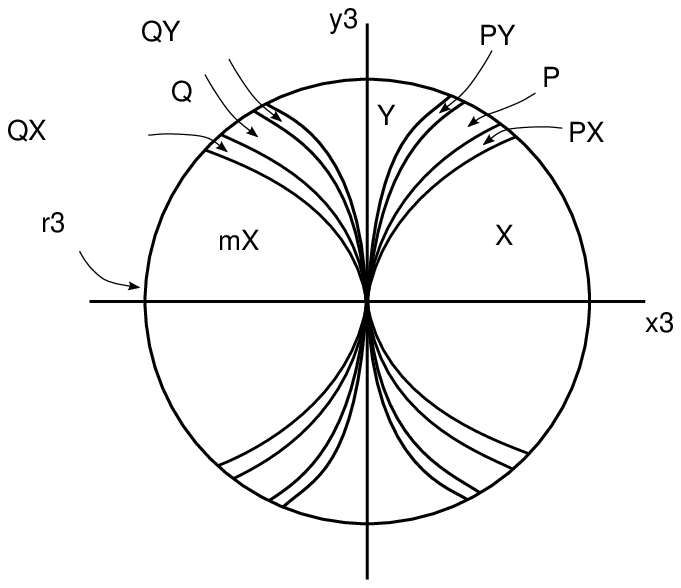} }
\subfigure[]{\includegraphics[width=8.7cm]{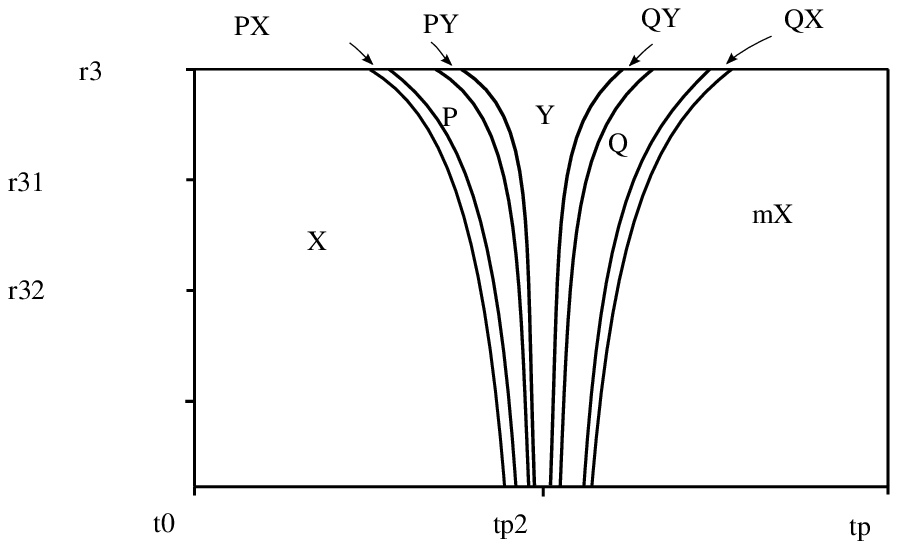} }
\end{center}
\caption{Schematic diagram showing a slice (with constant $u_2$) of the Poincar\'e section $\HBin$, using (a) Cartesian
and (b) logarithmic polar coordinates. Only part of the slice in (a) is shown in (b).
Each slice is divided into
regions according to the equilibrium in $C$ visited by the trajectories in that
region as they
pass from $B$ to~$A$.
}
 \label{fig:cusps}
 \end{figure}
 %%%%%%%%%%%%%%%%%%%%

Figure~\ref{fig:cusps} shows schematically how trajectories visit
different parts of~$C$, according to where they cross~$\HBin$. The majority of
trajectories go to $X$ or $-X$, and there are cusp-shaped regions that visit
$P$ then~$X$, $P$~only, etc., on their way to~$A$.

\begin{figure}
\psfrag{t0}{$\theta_3=0$}
\psfrag{tp2}{$\theta_3=2\pi$}
\psfrag{tp}{$\theta_3=\pi$}
\psfrag{ra}{$r_3=r_3^a$}
\psfrag{rb}{$r_3=r_3^b$}
\begin{center}
\includegraphics[width=12cm]{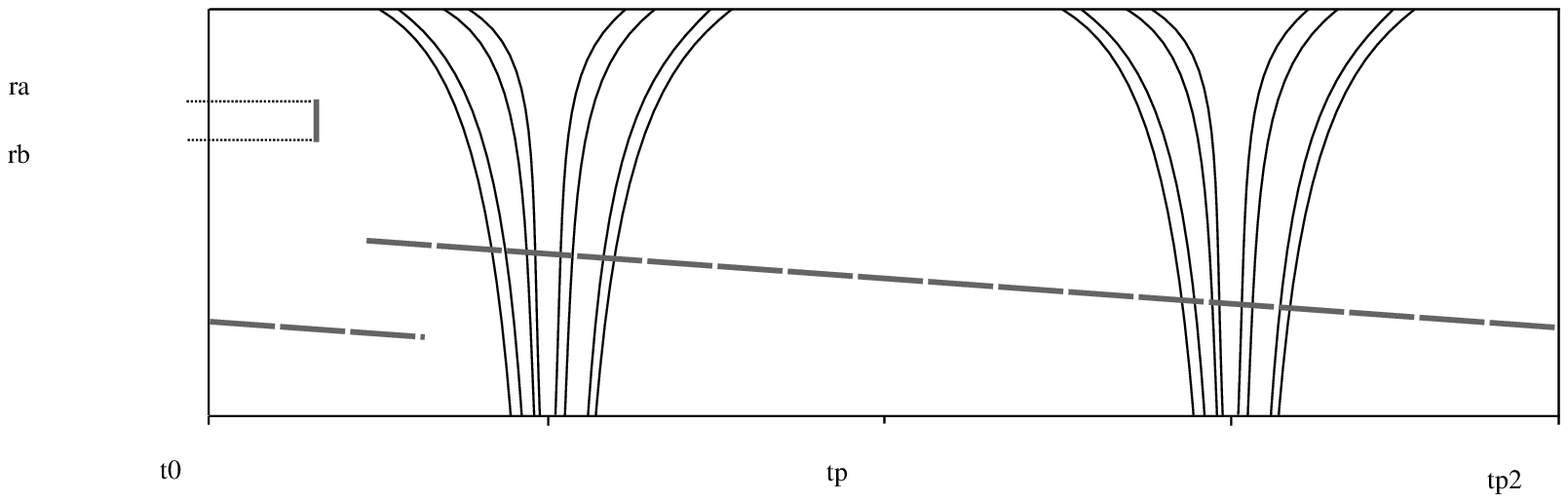}
\end{center}
\caption{Schematic diagram showing a line segment (solid grey line) on $\HBin$ (with
$0<r_3^a\leq r_3\leq r_3^b<h$, shown in
logarithmic polar coordinates) and its image (dashed grey line), after one
cycle around the network. The length of the line segment is chosen such that
its image covers the full range of values of~$\theta_3$. Note that in any
particular example, the image could be considerably more complicated than a straight line.}
 \label{fig:lineseg}
 \end{figure}

Consider a line segment of initial conditions on $\HBin$, with a fixed
value of $\theta_3$ and a range of values of $r_3$: $0<r_3^a\leq r_3\leq
r_3^b<h$ (see Figure~\ref{fig:lineseg}). We choose  $\theta_3$ such
that the family of trajectories first travels around the network via the point~$X$.
We will show that the spread of~$r_3$ values translates into a spread of~$\theta_3$  values
once the line segment has been mapped around the network. This arises in
particular from the complex eigenvalues at~$A$. We can choose the values of
$r_3^a$ and~$r_3^b$ such that after one cycle around the network, the line
segment covers $\theta_3\in[0,2\pi]$. This means that there are trajectories
with initial conditions in the initial segment
that visit each of the
different parts of~$C$ on the second
time around the network.

In order to show this, we repeat part of the calculation of
section~\ref{sec:stability} but starting on~$\HBin$ instead of~$\HAin$. For any
given initial value of~$r_3$ on $\HBin$, using inequalities (\ref{ineq:x1C}) and
(\ref{ineq:x2A}), the trajectory crosses~$\HAin$ with a value of~$\xiiAin$ that
satisfies:
 \begin{equation}
 \label{ineq:x2Ain}
   h\DCAmin
   \left(\frac{r_3}{h}\right)^{\deltaBmax\deltaCmax}
 \leq
 \xiiAin (r_3)
 \leq
 h\DCAmax
   \left(\frac{r_3}{h}\right)^{\deltaBmin\deltaCmin},
 \end{equation}
where
$\DCAmin=\DAmin\left(\DCmin\right)^{\deltaCmax}$ and
$\DCAmax=\DAmax\left(\DCmax\right)^{\deltaCmin}$.
On~$\HAin$, the trajectory has a value of~$\thetaAin$ close to that with which
the unstable manifold of~$X$ crosses~$\HAin$, since we chose our initial line
segment such that trajectories visited~$X$. This is essentially a constant,
but the complex eigenvalues mean that $\thetaAout$ depends logarithmically
on~$\xiiAin$. We want to choose $r_3^a$ and $r_3^b$ such that even allowing for
the range of values of $\xiiAin$ in~(\ref{ineq:x2Ain}), the local
map~(\ref{phiA}) guarantees that the range of values of $\thetaAout$ covers at
least $[0,2\pi]$. This requires that
 $$
 \frac{\omega}{\eA}
 \log\left(
     \frac{\min\left(\xiiAin(r_3^b)\right)}
          {\max\left(\xiiAin(r_3^a)\right)}
     \right)
 > 2\pi,
 $$
 where the minimum and maximum in this expression are taken over all possible
 values of $\theta_3$ on $\HBin$.
This last expression can be rewritten as
 $$
 \log\left(
     \frac{(r_3^b/h)^{\deltamax}}
          {(r_3^a/h)^{\deltamin}}
     \right)
 >
 \frac{2\pi\cA}{\omega}
 +
 \deltaA
     \log\left(
     \frac{\DCAmax}
          {\DCAmin}
     \right).
 $$
Values of $r_3^a$ and $r_3^b$ for which this is satisfied can clearly be found,
regardless of the values of the global constants or the values of~$\deltamin$
and~$\deltamax$, provided that $\omega\neq0$. Disregarding the global map
constants ($\DCAmax$~\hbox{etc.}), the inequality is satisfied if we choose
$r_3^b>r_3^a\exp\left(2\pi\cA/\omega\deltamax\right)$. By choosing $r_3^a$
small enough (close enough to the network), the length of the initial line
segment $r_3^b-r_3^a$ can be made as small as we wish.

With this choice of $r_3^a$ and $r_3^b$, the line segment of initial conditions
maps to at least a full circle on~$\HAout$ and consequently, using the global
map~(\ref{PsiAB}), also on~$\HBin$, since that map rigidly rotates the angle.
Therefore, on their next circuit around the network, trajectories from within
this family could visit any of $\pm{X}$, $\pm{Y}$, $\pm{P}$ or~$\pm{Q}$.

The same argument holds with minor changes regardless of the location of the
initial line segment, so we conclude that close to the network, there are
trajectories that visit any equilibrium point in~$C$ followed by any other equilibrium point in~$C$ on
two consecutive circuits of the network. Since there is freedom in choosing
the exact location of the line segment, this argument implies the same
conclusion can be drawn for open sets of initial conditions. Thus we have shown
that arbitrarily close to the network, there are open sets of orbits that
switch from any route around the network to any other route.

The argument above does not require the network to be asymptotically stable, and
only refers to two consecutive circuits of the network, and so does not
demonstrate that typical trajectories will continue to switch for ever as they
evolve. In the remainder of this section we consider long term switching, first in the
case where the network is attracting ($\deltamin>1$) and then in the case
where $\deltamin<1<\deltamax$.

The argument above implies that of trajectories starting in a typical ball of initial
conditions close to the network and first visiting~$X$, fewer than
half will go on to visit~$X$ on their second circuit of the
network. In the case $\deltamin>1$, all trajectories starting in a typical ball approach the network,
so their values of $\xiiAin$ get smaller and smaller, and hence the values of
$\thetaAout$ for these trajectories get more and more spread out. Therefore, the argument that the
chance of visiting $X$ on the next cycle around the network is less than half
continues to hold; in the limit, we expect that the set of trajectories that
visit only $X$ and never visit $-X$ has measure zero. Thus, typical
trajectories should visit both~$\pm{X}$ roughly equally, and there is always a
chance they could visit $\pm{Y}$, $\pm{P}$ or $\pm{Q}$ as well.

We can estimate the probabilities of visiting different parts of the network by
computing the proportion of trajectories starting on $\HBin$ that will visit $X$
or $Y$ on their next time around the network. Consider a circle of initial
conditions on $\HBin$ with $r_3=a$ ($a<h$) and $0\leq\theta_3<2\pi$. Of those
in the first quadrant, some will visit~$X$, some will visit $P$ and the rest
will visit~$Y$ (see Figure~\ref{fig:cusps}). The boundaries between the
different possibilities are of the form $x_3=Ky_3^{\alpha}$, where
$\alpha=\eBx/\eBy>1$, and $K$~is a constant that depends on which
boundary is being considered. To simplify the discussion, we omit the details
of those trajectories that visit~$P$, and consider only a single boundary that
separates trajectories that go via $X$ and those that go via~$Y$.

The intersection of the boundary with the circle $r_3=a$ can be found by
solving
 $$
 a^2 = x_3^2 + y_3^2
     = y_3^2\left(1 + K^2y_3^{2\alpha-2}\right)
 $$
for $y_3$ as a function of $a$, $\alpha$ and~$K$. For small~$a$ and for
$\alpha>1$, the solution is approximately
 $$
 y_3 = a\left(1 - \frac{1}{2}K^2a^{2\alpha-2}\right),
 \qquad
 x_3 = Ka^{\alpha}.
 $$
The proportion of trajectories starting on the circle $r_3=a$ in $\HBin$ that
visit~$Y$ is approximately equal to $2x_3/2\pi{a}=Ka^{\alpha-1}/\pi$ for small~$x_3$.
The same proportion visit~$-Y$,
and the remainder are split equally between $X$ and~$-X$.

In the case that the network is attracting ($\deltamin>1$), the value of $r_3$
on~$\HBin$ decreases each time around the network, so $a\rightarrow0$.
Trajectories spend increasingly long periods of time near~$A$, where the
eigenvalues are complex, so one would expect that the angle~$\theta_3$ becomes
essentially a random variable. In this case, the chance of visiting~${Y}$ or ${-Y}$
goes to zero, and the chances of visiting $X$ or~$-X$ will both tend
to~$\frac{1}{2}$.

In the case $\deltamax<1$, where the network is unstable, trajectories leave
the neighbourhood of the network and no estimates are possible.

The intermediate case ($\deltamin<1<\deltamax$) offers the interesting
possibility
that trajectories might maintain an average distance from the network, either
in a periodic or chaotic fashion. In the latter case, one might expect that an
average, weighted using the probabilities above, of the contraction around one
part of the network and the expansion around the other part might lead to
conditions for the existence of a nearby invariant set. Making a weighted
average in this way only makes sense if trajectories switch \emph{irregularly}
between $\pm{X}$ and~$\pm{Y}$, so an existence condition could include a
requirement for switching. In the case of a periodic orbit, the weighting would
depend on the itinerary of the orbit. This weighted average would also depend
on~$a$, which we interpret as the average distance from the network.
We give numerical examples of this
phenomenon below, and defer a detailed analysis of this case to a later paper.

\section{Numerical example}
\label{sec:numeg}

In this section we present some numerical results based on the following
equations:
\begin{equation}
\label{numex}
\begin{array}{lcl}
\dot{x}_1 &=& x_1(1-x_1^2-Ex_2^2), \\
\dot{x}_2 &=& x_2(1-x_2^2-Fx_3^2-Gy_3^2), \\
\dot{x}_3 &=& x_3(1-Ex_1^2+Hx_2^2-x_3^2-Dy_3^2) - \omega y_3x_1^2, \\
\dot{y}_3 &=& y_3(1-Ex_1^2-Dx_3^2-y_3^2) + \omega x_3x_1^2, \\
\end{array}
\end{equation}
where $D$, $E$, $F$, $G$ and $H$ are parameters that we vary in our numerics.
The parameter~$H$ controls the relative
values of the two expanding eigenvalues at the point~$B$. Throughout we assume
that $H>0$, and that  $D, E, F, G, (F+G)/(D+1)\in(1,3)$, so that assumptions {\bf A1}--{\bf A8} from
section~\ref{sec:hetnet} are satisfied.
The values of all parameters and eigenvalue ratios are given in Table~\ref{tab:eigenvalues}.

\begin{table}
\begin{center}
\begin{tabular}{|c|c|c|}
\hline
Equilibrium point & Eigenvalues & Eigenvalue ratios \\
\hline
  $A$: $(1,0,0,0)$
  & $\rA=2$, $\eA=1$, $\cA=E-1$
  & $\delta_A=E-1$ \\
\hline
  $B$: $(0,1,0,0)$
  & $\rB=2$, $\cB=E-1$,
  & $\deltaBx=\frac{E-1}{1+H}$ \\
  & $\eBx=1+H$, $\eBy=1$
  & $\deltaBy=E-1$ \\
\hline
  $X$: $(0,0,1,0)$
  & $\rC(\theta_3^X)=2$, $\lambda_X=D-1$,
  & $\deltaCX=F-1$ \\
  & $\eC(\theta_3^X)=1$, $\cC(\theta_3^X)=F-1$
  & \\
\hline
  $Y$: $(0,0,0,1)$
  & $\rC(\theta_3^Y)=2$, $\lambda_Y=D-1$,
  & $\deltaCY=G-1$ \\
  & $\eC(\theta_3^Y)=1$, $\cC(\theta_3^Y)=G-1$
  & \\
\hline
  $P$: $(0,0,{1\over \sqrt{D+1}},{1\over\sqrt{D+1}})$
  & $\rC(\theta_3^P)=2$, $\lambda_P=2\frac{D-1}{D+1}$,
  & $\deltaCP=\frac{F+G}{D+1}-1$ \\
  & $\eC(\theta_3^P)=1$, $\cC(\theta_3^P)=\frac{F+G}{D+1}-1$
  & \\
\hline
  $Q$: $(0,0,-{1\over \sqrt{D+1}},{1\over\sqrt{D+1}})$
  & $\rC(\theta_3^Q)=2$, $\lambda_Q=2\frac{D-1}{D+1}$,
  & $\deltaCQ=\frac{F+G}{D+1}-1$ \\
  & $\eC(\theta_3^Q)=1$, $\cC(\theta_3^Q)=\frac{F+G}{D+1}-1$
  & \\
\hline
\end{tabular}
\vspace{1ex}
\caption{Eigenvalues associated with equilibria of
equations~(\ref{numex}), and values of eigenvalue ratios.
The actual eigenvalues at~$A$ (for example) are $-\rA$
(radial), $-\cA\pm{i}\omega$ (contracting) and $\eA$ (expanding).
The eigenvalues $-\lambda_X$, $\lambda_P$, $-\lambda_Y$ and
$\lambda_Q$ all have corresponding eigenvectors tangent to the invariant circle~$C$.}
\label{tab:eigenvalues}
\end{center}
\end{table}

\begin{table}
\begin{center}
\begin{tabular}{|c|c|c|c|}
\hline
Parameter & I & II & III \\
\hline
  $F$
  & 1.63125
  & 1.63125
  & 1.61875 \\
\hline
  $G$
  & 1.671386719
  & 1.549316406
  & 1.671386719 \\
\hline
\hline
  $\deltaBx=\deltaBmin$
  & 1.25
  & 1.25
  & 1.25 \\
\hline
  $\deltaBy=\deltaBmax$
  & 1.28
  & 1.28
  & 1.28 \\
\hline
  $\deltaCX$
  & 0.63125
  & 0.63125
  & 0.61875 \\
\hline
  $\deltaCY$
  & 0.671386719
  & 0.549316406
  & 0.671386719 \\
\hline
  $\deltaCP=\deltaCQ$
  & 0.63496867
  & 0.57453782
  & 0.62878055 \\
\hline
  $\deltaX$
  & 1.01
  & 1.01
  & 0.99 \\
\hline
  $\deltaY$
  & 1.10
  & 0.90
  & 1.10 \\
\hline
  $\deltamax$
  & 1.10
  & 1.0342400
  & 1.10 \\
\hline
  $\deltamin$
  & 1.01
  & 0.87890620
  & 0.99 \\
\hline
\end{tabular}
\vspace{1ex}
\caption{Values of the parameters and the eigenvalue ratios
for the three examples. The values in common are $\omega=1$, $H=0.024$, $D=1.02$ and
$E=2.28$. The parameters $F$ and $G$ are set using $F=1+\deltaX(1+H)/(E-1)^2$
and $G=1+\deltaY/(E-1)^2$.}
\label{tab:calculations}
\end{center}
\end{table}

To simplify the presentation, we always choose the parameter~$D$ so that
$\deltaCP$ and $\deltaCQ$ are intermediate between $\deltaCX$ and $\deltaCY$.
With this constraint, the
additional combinations of eigenvalue ratios are
 $\deltaCmax=\max\left(\deltaCX,\deltaCY\right)$,
 $\deltaCmin=\min\left(\deltaCX,\deltaCY\right)$. Recall that
 $$
 \deltaBmax=\deltaBy,
 \quad
 \deltaBmin=\deltaBx,
 \quad
 \deltamax=\deltaA\deltaBmax\deltaCmax,
 \quad
 \deltamin=\deltaA\deltaBmin\deltaCmin.
 $$
We make the further definitions
 $\delta_X=\deltaA\deltaBx\deltaCX$ and
 $\delta_Y=\deltaA\deltaBy\deltaCY$\/: these are the eigenvalue ratios relevant
to trajectories that leave~$B$ along the $x_3$ (resp.,~$y_3$) direction and
visit~$\pm{X}$ (resp.,~$\pm{Y}$).

To illustrate typical dynamics near the network, we present three examples.
In these examples, we choose values of~$\delta$ close to~$1$
so that trajectories do not approach or leave the network too quickly.
The first example has $\deltamin>1$, and the other two have
$\deltamin<1<\deltamax$, with different choices as to whether it is $\deltaX$ or $\deltaY$
that is less than~$1$. The specific choices of parameters and eigenvalue ratios for the different
examples are given
in Table~\ref{tab:calculations}.
 \begin{itemize}
 \item Example~I:  $\deltamin=\deltaX=1.01$ and
          $\deltamax=\deltaY=1.10$. In this example, trajectories approach the
network, predominantly switching between~${X}$ and $-X$.
 \item Example~II:  $\deltamin<\deltaY=0.90$ and
           $\deltamax>\deltaX=1.01$. In this case trajectories behave much as
in Example~I, even though the network is not asymptotically stable, since
almost all trajectories visit~$\pm{X}$ most of the time and so are most heavily
influenced by the value of~$\deltaX$.
\item Example~III:  $\deltamin=\deltaX=0.99$ and
            $\deltamax=\deltaY=1.10$. Here trajectories leave a neighbourhood of the
network and end up displaying periodic or chaotic switching.
\end{itemize}

Care needs to be taken with numerical integration of systems with heteroclinic cycles and
networks, because of the potential for rounding errors to cause qualitatively incorrect
results.
We first integrated equations~(\ref{numex}) numerically using the Bulirsch--Stoer
adaptive integrator~\cite{Press1986}, with a tolerance for the relative error
set to $10^{-12}$ for each step.  We also rewrote
equations~(\ref{numex}) using logarithmic variables $(\log x_1, \log x_2,
\log r_3, \theta_3)$ instead of $(x_1,x_2,x_3,y_3)$, and integrated the
converted equations with a tolerance of~$10^{-10}$, rising to $10^{-8}$ for
trajectories very close to the network. This enabled us to examine whether the
numerical methods handle the very large dynamic range of the variables without
being unduly affected by rounding errors. The two methods of computing
solutions agree to within the specified tolerance when we compute periodic
trajectories (in calculations~II and~III), and they agree to within the
specified tolerance for times up to about $3000$ when we compute trajectories
very close to the network. Beyond this time, trajectories computed by the two methods
diverge, but the qualitative behaviour of the trajectories is the same.  We
have confirmed that the results are not sensitive to the exact value of the the
relative error tolerance that we chose. The results shown in the figures below
are all computed using logarithmic variables.

Poincar\'e sections were
computed using algorithms from~\cite{Parker1989}. The nodes on the network are all
simple equilibria lying within coordinate planes, so the numerical issues
associated with cycling chaos (chaotic dynamics within nodes on the network),
as discussed for example in~\cite{Ashwin2003e,Ashwin2004} do not arise here.

%%%%%%%%%%%%%
\begin{figure}
%\psfrag{x3title}{$x_3$}
%\psfrag{y3title}{$y_3$}
%\psfrag{pxtitle}{$X$}
%\psfrag{mxtitle}{$-X$}
%\psfrag{pytitle}{$Y$}
%\psfrag{mytitle}{$-Y$}
%\psfrag{pptitle}{$P$}
%\psfrag{mptitle}{$-P$}
%\psfrag{pqtitle}{$Q$}
%\psfrag{mqtitle}{$-Q$}
\begin{center}
\includegraphics[width=8cm]{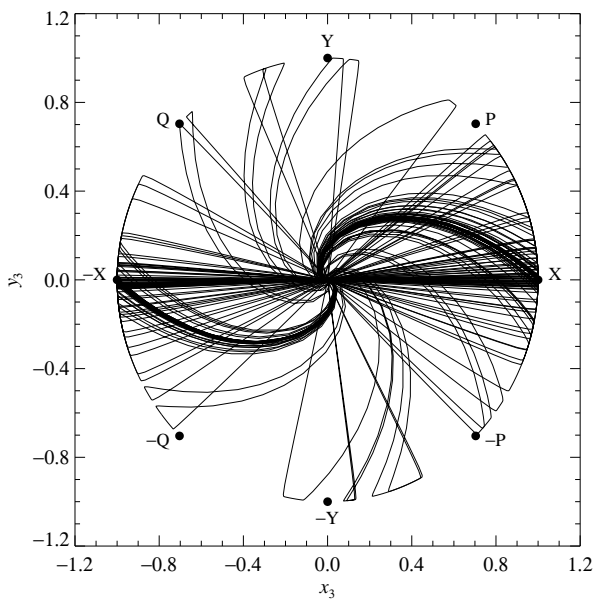}
\end{center}
\caption{Example~I: Phase portrait showing the $(x_3,y_3)$ projection of a single
trajectory.
The trajectory leaves~$B$ (at the
origin in this projection), goes to~$C$ (approximately the
circle $x_3^2+y_3^2=1$) in a more-or-less straight line,
travels around~$C$ towards~$\pm{X}$ or~$\pm{Y}$,
then spirals in
to~$A$ (also at the origin) before returning to~$B$.
}
 \label{fig:calciphase}
 \end{figure}
 %%%%%%%%%%%%

 %%%%%%%%%%%%%%
\begin{figure}
%\psfrag{x1title}{$x_1$}
%\psfrag{x2title}{$x_2$}
%\psfrag{x3title}{$x_3$}
%\psfrag{y3title}{$y_3$}
%\psfrag{timetitle}{$t$}
\begin{center}
\includegraphics[width=\hsize]{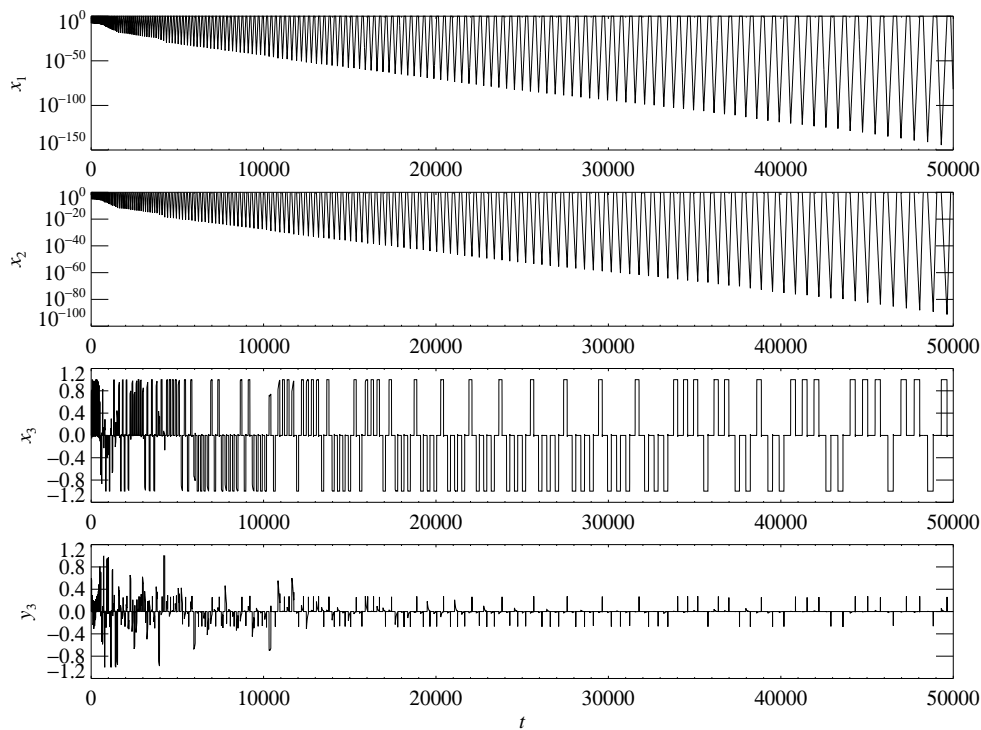}
\end{center}
\caption{Example~I: Time series for the trajectory shown in
Figure~\ref{fig:calciphase}. The $x_1$ and $x_2$ plots show that the
trajectory is approaching the network. The $x_3$ plot shows the repeated
switching between~${X}$ and $-X$, and the $y_3$ plot shows that visits to~$\pm{Y}$
become increasingly rare as the trajectory gets closer to the network.}
\label{fig:calcitimelong}
\end{figure}
%%%%%%%%%%%%%%%%%

\subsection{Example~I: Trajectories approach the stable network}

In this example, $\deltamin=\deltaX=1.01$ and $\deltamax=\deltaY=1.10$ are both
greater than one, and the network is asymptotically stable.
The phase portrait and time series shown in Figures~\ref{fig:calciphase}  and~\ref{fig:calcitimelong}
correspond to a trajectory started
from the initial condition $x_1(0)=0.01=h$, $x_2(0)=1$ and $x_3(0)=y_3(0)=10^{-5}$,
and illustrate the occurrence of repeated switching in the transient
dynamics.
As expected from section~\ref{sec:switching}, a typical trajectory lying near
the network makes repeated switches between ${X}$ and $-X$ and visits $\pm{Y}$
occasionally, but the visits to~$\pm{Y}$ become increasingly
rare as the trajectory gets closer to the network.

%%%%%%%%%%%%
\begin{figure}
%\psfrag{r3title}{$r_3$}
%\psfrag{thetatitle}{$\theta_3/2\pi$}
\begin{center}
\includegraphics[width=\hsize]{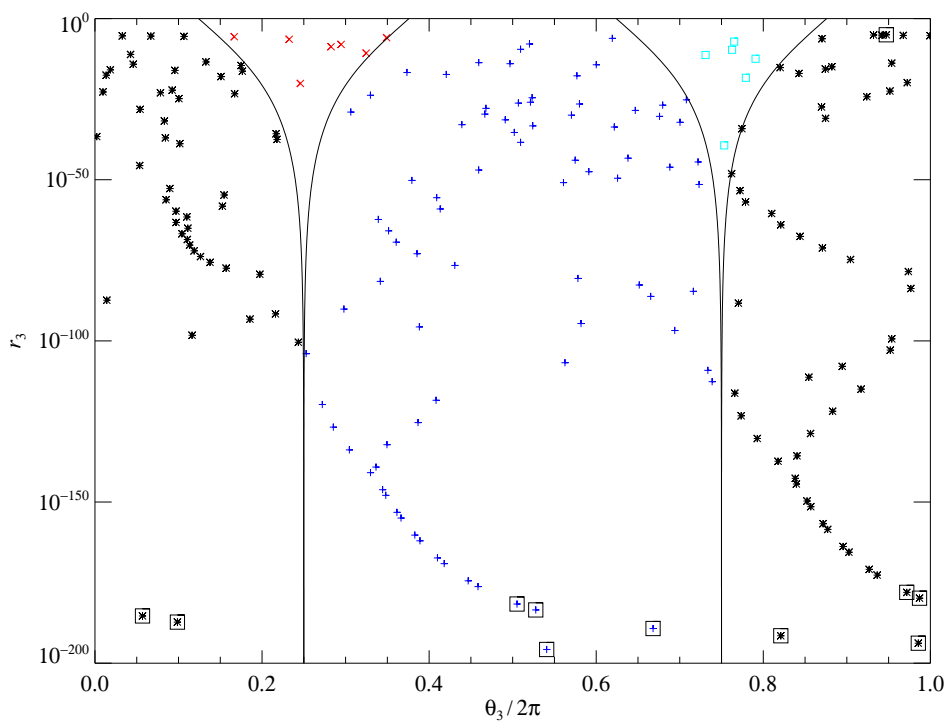}
\end{center}
\caption{Example~I: Projection of a Poincar\'e section for the trajectory shown in
Figure~\ref{fig:calciphase}. Black asterixes (blue plusses) indicate that the
trajectory visits
$X$ ($-X$) immediately after leaving the Poincar\'e section; red crosses
(cyan boxes) indicate that the trajectory next visits $Y$ ($-Y$). The
single point in a box at the top is the first point in the trajectory, and the
ten boxed points at the bottom are the final points. This figure illustrates how the
trajectory approaches the network; the trajectory initially visits $\pm{X}$ and
$\pm{Y}$, but as it approaches the network, visits to $\pm{Y}$ become rare while
switching between ${X}$ and $-X$ is persistent. The boundaries of the cuspoidal
regions are indicative of the boundaries between trajectories that have different
routes on their next circuit of the network (c.f., Figure~\ref{fig:cusps}).
They have been chosen to match the available data for
larger~$r_3$.}
 \label{fig:calcipoincare}
 \end{figure}
 %%%%%%%%%%%%%%

Figure~\ref{fig:calcipoincare} shows where the trajectory intersects the
Poincar\'e section~$\HBin$, defined here as $x_1=h=0.01$, $x_2\approx1$ and $r_3<{h}$.
After leaving~$B$, the trajectory visits~$\pm{X}$ or~$\pm{Y}$: the symbols indicate
which of the four possibilities occurs immediately after the
intersection marked. The boundaries separating regions of $\HBin$ from which trajectories
leave for $\pm{X}$
and~$\pm{Y}$ can be clearly seen, and are consistent with the results sketched in Figure~\ref{fig:lineseg}. As
the trajectory approaches the network, travelling from top to bottom in
Figure~\ref{fig:calcipoincare}, visits to~$\pm{Y}$ are not seen for
$r_3<10^{-20}$ or so, although they are in principle possible for arbitrarily
small~$r_3$. Visits to $X$ and $-X$
do not occur in a periodic fashion.

\begin{figure}
%\psfrag{r3title}{$r_3$}
%\psfrag{thetatitle}{$\theta_3/2\pi$}
\begin{center}
\includegraphics[width=\hsize]{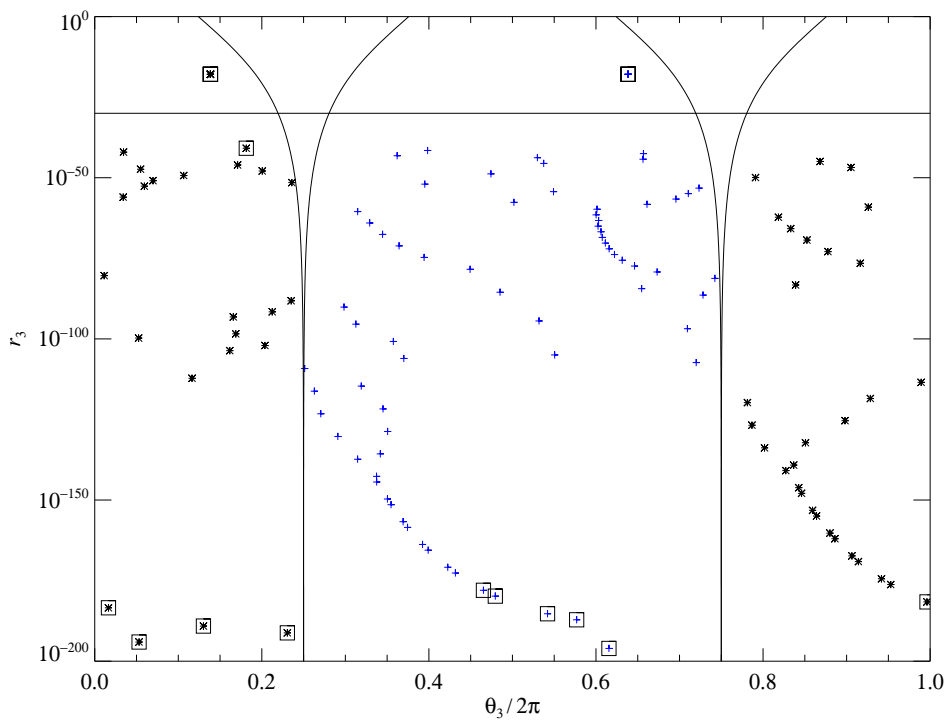}
\end{center}
\caption{Example~II: Poincar\'e section.
The trajectory with initial condition $x_1(0)=0.01=h$, $x_2(0)=1$, $x_3(0)=y_3(0)=10^{-40}$
(below the horizontal
line) approaches the network, starting at the single boxed point just
below the line, and ending at the ten boxed points at the bottom of the figure.
The trajectory with initial condition $x_1(0)=0.01=h$, $x_2(0)=1$, $x_3(0)=y_3(0)=10^{-20}$
approaches a
stable periodic orbit represented by the two boxed points above the line.
Symbols are as in Figure~\ref{fig:calcipoincare}.}
\label{fig:calciipoincare}
\end{figure}

\subsection{Example~II: Trajectories approach the unstable network}

In this example, $\deltamin<\deltaY=0.90$ and
$\deltamax>\deltaX=1.01$, and the network is asymptotically unstable.
However, trajectories that start close enough to the network
can still approach the network. For instance, the initial condition
$x_1(0)=0.01=h$, $x_2(0)=1$, $x_3(0)=y_3(0)=10^{-40}$ yields the trajectory shown in
Figure~\ref{fig:calciipoincare} (below the horizontal line). This trajectory
never visits~$\pm{Y}$ but
does switch repeatedly between~${X}$ and $-X$ while getting closer to the network. In
contrast, the initial condition $x_1(0)=0.01=h$, $x_2(0)=1$, $x_3(0)=y_3(0)=10^{-20}$
yields the periodic orbit close to the network shown above the horizontal
line in Figure~\ref{fig:calciipoincare}.
There are other stable periodic orbits further away from the network.

This behaviour is consistent with the discussion in section~\ref{sec:analysis}.
Since~$\deltaX>1$, a trajectory that starts close enough to the network will
mostly only visit~$\pm{X}$ (almost never $\pm Y$) and so can approach the network even if $\deltaY<1$. On the other hand, the discussion in section~\ref{sec:analysis} predicts that there are
trajectories arbitrarily close to the
network that visit~$\pm{Y}$ sufficiently often to be repelled from the network,
and so the network is unstable. We have found examples of stable
periodic orbits that are quite close to the network; most likely there are
unstable periodic orbits as well, but we have not explored this possibility.
We conjecture that the measure of initial conditions that do not eventually go
to the network tends to zero as these get closer to the network, so the network
will be essentially asymptotically stable. The reason for this is that the
cusps delimiting trajectories that go to~$\pm{Y}$ (and all their preimages) are
thin.

\begin{figure}
%\psfrag{r3title}{$r_3$}
%\psfrag{thetatitle}{$\theta_3/2\pi$}
\begin{center}
\includegraphics[width=\hsize]{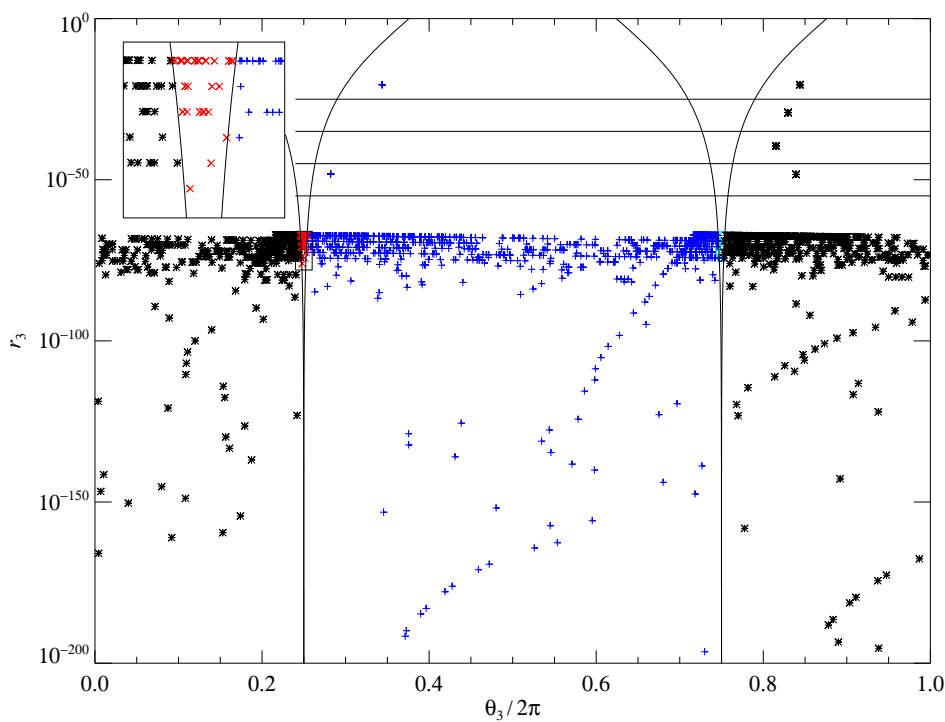}
\end{center}
\caption{Example~III: Poincar\'e section, showing five trajectories
separated by horizontal lines. With $x_1(0)=0.01=h$, $x_2(0)=1$,
$x_3(0)=y_3(0)=10^{-200}$, the
trajectory starts at the bottom of the figure and moves away from the network, but finds a
chaotic attractor with $r_3<10^{-67}$ (below the lowest line). This
trajectory chaotically switches between~$\pm{Y}$ and~$\pm{X}$.
Above this, there are
four examples of stable periodic orbits.
The inset enlarges the boxed region near $\theta_3=\pi/2$, $r_3 = 10^{-70}$.
Symbols and colours are as in Figure~\ref{fig:calcipoincare}.}
\label{fig:calciiipoincare}
\end{figure}

\subsection{Example~III: Trajectories leave the unstable network}

The final example has $\deltamin=\deltaX=0.99$ and $\deltamax=\deltaY=1.10$,
and shows that although the network is unstable, there are nearby periodic and
chaotic orbits. The Poincar\'e section in Figure~\ref{fig:calciiipoincare}
shows five trajectories. Each initially has $x_1(0)=0.01=h$ and $x_2(0)=1$, and the
$x_3$ and $y_3$ initial conditions vary between trajectories: from bottom to top in the
figure, the initial conditions are $x_3(0)=y_3(0)=10^{-200}$,
$10^{-50}$, $10^{-40}$, $10^{-30}$ and $10^{-20}$, with the resulting trajectories
being separated by horizontal
lines in the figure. Trajectories starting with the four largest initial conditions find
stable periodic orbits; there are most likely additional orbits that we have not
found. In contrast, trajectories starting very close to the network move away
and find a chaotic attractor at around $r_3\approx10^{-70}$. Within this
attractor, the trajectory mostly visits~$\pm{X}$ and moves gradually away from
the network, since~$\deltaX<1$. As it does so, the chance of visiting~$\pm{Y}$
increases; when this happens (with~$\deltaY>1$), the trajectory jumps closer to the
network.
The same chaotic attractor was found with several initial conditions
in the range $10^{-200}\leq{}x_3(0)=y_3(0)\leq10^{-60}$. This  example of
sustained chaotic switching between~$\pm{X}$ and~$\pm{Y}$ illustrates the
ideas discussed in section~\ref{sec:switching}.

\section{Conclusions}
\label{sec:conc}

In this paper we have illustrated a simple mechanism that produces
switching between different structurally stable
heteroclinic cycles in a heteroclinic
network, namely the presence of complex eigenvalues in the linearisation
about one of the equilibria common to all cycles in the network.
This is done in the context of an example in ${\mathbb R}^4$
with ${\mathbb Z}_2^3$ symmetry. Switching arising from the
presence of complex eigenvalues has been seen in other examples
\cite{Aguiar2005,Aguiar2009a} but in those cases the cycles are structurally stable because of
transversal intersections of some manifolds rather than purely because of
the presence of symmetry.

By the construction and analysis of maps that model the
dynamics near cycles in our network, we have
found a simple condition under which the heteroclinic
network is asymptotically stable. The construction of the maps
used standard techniques that were modified to allow us to keep track
of the continuum of heteroclinic cycles present in our network.
A crucial step in the analysis of the network was recognition that
the network could efficiently be treated as a collection of connections
between equilibria ($A$ and $B$) and an invariant circle ($C$)
rather than a collection of cycles each of which connected a set of
equilibria. Earlier attempts to treat the heteroclinic connections going to each of
the equilibria on $C$ (i.e., to $\pm X$, $\pm Y$,$ \pm P$ and $\pm Q$) separately
proved to be intractable and were ultimately unfruitful.

We found that the network is asymptotically stable if $\deltamin>1$.
The quantity $\deltamin$ was defined in section \ref{sec:stability}
and is the product of the ratios of the (real part of the) contracting and expanding
eigenvalues at selected equilibria
in the network, namely at $A$, $B$, and at the equilibrium on $C$ at which the
ratio of contracting to expanding eigenvalues is minimised. At $B$ there are two
expanding eigenvalues, and $\deltamin$ is defined using the larger of
these two eigenvalues. Effectively, $\deltamin$ is the minimum ratio of
contracting to expanding eigenvalues that could be encountered by a trajectory
on one circuit through the network, starting and finishing at one of the common
equilibria $A$ or $B$.

Another important quantity for network stability is $\deltamax$. As defined
in section \ref{sec:stability}, $\deltamax$ is effectively the maximum ratio
of contracting to expanding eigenvalues that could be encountered by a trajectory
on one circuit through the network. We showed that if $\deltamax<1$, then a
trajectory started close to the network (but not on the stable manifold of any of the
equilibria) will be further away from the network
after one circuit of the network, regardless of its itinerary, and thus the
network is unstable.

These results on network stability are a natural generalisation
of established stability results for heteroclinic cycles, where it has been shown
that asymptotic stability of a cycle is often determined (in part, at least) by
the ratio of contracting to expanding eigenvalues along the cycle.

We have shown
that switching is ubiquitous near our network. In particular,
we showed that close enough to the network, there are
trajectories that, over the course of two cycles around the network, visit any
combination of the equilibrium points within~$C$ in any order.
(A similar result holds for the examples of~\cite{Aguiar2005,Aguiar2009a}.)
This occurs regardless of whether or not the
network is asymptotically stable. In the case that the
network is asymptotically stable, we showed that most trajectories repeatedly visit both~$X$
and $-X$ as they approach the network, while, on the assumption that the
complex eigenvalues at~$A$ mix trajectories effectively,
visits to~$\pm{Y}$ become rare.
Additive noise could clearly have an important effect on the switching
behaviour; we have not explored this issue.

Our results about repeated switching in the network and our
categorization of network stability  in terms of $\deltamin$ and $\deltamax$
has allowed us to
identify an interesting case, intermediate between asymptotic stability and complete instability of the network.
Specifically, if $\deltamin < 1 < \deltamax$ then whether or not an individual
trajectory approaches the network or diverges from it depends on the detailed itinerary of that
trajectory. We found two main cases. First, if the ratio of contracting to expanding eigenvalues
encountered by a trajectory making a circuit near the dominant cycle in the network (i.e., a cycle
involving a visit to either $X$ or $-X$) is greater than one ($\delta_X>1$, using the notation of
section \ref{sec:numeg}), then we conjecture that almost all trajectories will
eventually converge to the network, even though the network is asymptotically unstable.
If on the other hand, the dominant cycle has the appropriate ratio of eigenvalues less than one
($\delta_X<1$) with one of the other cycles having a ratio greater than one ($\delta_Y>1$, in the notation
of section \ref{sec:numeg}) then there might be a delicate balance between the repelling properties
of the dominant cycle and the attracting properties of the other cycle; most trajectories would
be repelled from the unstable network but may be attracted to a chaotic or periodic attractor some
small distance away from the network. In this case, the network structure may still be observed in
the long term dynamics of the system even though the network is unstable.

An example of this phenomenon was shown
in section \ref{sec:numeg} where results from numerical integration of
a particular system of differential equations were presented.
We have not attempted to quantify the balance that occurs in the example of a
chaotic attractor in section~\ref{sec:numeg}, but defer this to a later paper.
We note that heteroclinic networks with delicate stability properties have
been studied before (e.g., in \cite{Kirk1994}); the point of difference
here is that the switching mechanism operating in our network ensures that
most trajectories will visit most parts of a neighbourhood of the network.
The transition from $\deltamin>1$ to $\deltamin<1$ is an example of a resonance
of the heteroclinic network, and it is clear from this example that the network
structure will make analysis of the resonance quite involved. We defer this
analysis to a future paper.

\section*{Acknowledgments}
This research has been supported
by the University of Auckland Research Council, the Engineering and Physical
Sciences Research Council (EP/G052603/1) and the National Science Foundation
(DMS-0709232). We are grateful for the
hospitality of the Department of Mathematics at the University of Auckland,
the Department of Engineering Sciences and Applied Mathematics at Northwestern University, and
the School of Mathematics at the University of Leeds.

\bibliographystyle{plain}
\bibliography{kprs}

\end{document}